\documentclass[preprint,showpacs,showkeys]{revtex4}

\usepackage{graphicx}
\usepackage{amsmath}
\usepackage{amssymb}
\usepackage{amsfonts}
\usepackage{bm}
\usepackage{subfigure}

\begin{document}
\bibliographystyle{apsrev}

\title{Two-dimensional Ising model with competing interactions and
its application to clusters and arrays of $\pi$-rings and
adiabatic quantum computing}

\author{A.~O'Hare$^1$, F.~V.~Kusmartsev$^1$, K.~I.~Kugel$^{1,2}$,
and M.~S.~Laad$^1$}

\affiliation{$^1$Department of Physics, Loughborough University,
Leicestershire, LE11 3TU, UK}

\affiliation{$^2$ Institute for Theoretical and Applied
Electrodynamics, Russian Academy of Sciences, Izhorskaya Str. 13,
Moscow, 125412 Russia}

 \date{\today}

\begin{abstract}
We study planar clusters consisting of loops including a Josephson
$\pi$-junction ($\pi$-rings). Each $\pi$-ring carries a persistent
current and behaves as a classical orbital moment. The type of
particular state associated with the orientation of orbital
moments at the cluster depends on the interaction between these
orbital moments and can be easily controlled, i.e. by a bias
current or by other means. We show that these systems can be
described by the two-dimensional Ising model with competing
nearest-neighbor and diagonal interactions and investigate the
phase diagram of this model. The characteristic features of the
model are analyzed based on the exact solutions for small clusters
such as a 5-site square plaquette as well as on a mean-field type
approach for the infinite square lattice of Ising spins. The
results are compared with spin patterns obtained by Monte Carlo
simulations for the 100 $\times$ 100 square lattice and with
experiment. We show that the $\pi$-ring clusters may be used as a
new type of superconducting memory elements. The obtained results
may be verified in experiments and are applicable to adiabatic
quantum computing where the states are switched adiabatically with
the slow change of coupling constants.
\end{abstract}


\smallskip

\maketitle

\section{Introduction.}

For several decades the formation of different kinds of
superstructures in solids remains  a topical issue in condensed
matter physics. The superstructures (or spatially modulated
structures) may be of a different nature: magnetic patterns like
spin-density waves, inhomogeneous charge distributions in
charge-ordered compounds, dipolar and quadrupolar ordering in
ferroelectrics or ferroelastics, regular lattice distortions and
related orbital structures, stripe-like arrangements of dopants in
alloys, etc. The phase diagrams of such compounds can be rather
complicated involving a large number of phases with non-trivial
types of ordering. Fortunately, all this wealth of seemingly
unrelated phenomena can be often described by rather simple models
with a due account taken of a competitive character of  the most
important interactions. One of the most popular models of such
type was proposed by Elliott \cite{Elliott} as early as in 1961,
who analyzed specific features of magnetic ordering in heavy
rare-earth metals. This model named by Fisher and Selke
\cite{FisherSelke} anisotropic next-nearest neighbor Ising (ANNNI)
model in its initial form describes a cubic lattice of Ising spins
composed of ferromagnetic planes with the nearest-neighbor
spin-spin interaction, whereas there exists the ferromagnetic
interaction $J_1$ between neighboring planes and the
antiferromagnetic interaction $J_2$ between next-nearest planes.
Owing to this rather straightforwardly introduced competition of
interactions, the ANNNI model exhibits an unexpectedly complicated
phase diagram in the $J_2/J_1-T$ plane being a manifestation of
the so called ``devil's staircase'' \cite{Bak86,Bak82}. The
properties of this model were thoroughly studied (see, e.g.
\cite{Bak80}) and it was successfully applied to the analysis of
numerous systems exhibiting modulated structures (for review, see
\cite{SelkePRep} and references therein). This work still
continues: as an example, we can mention recent paper
\cite{Kimura} where a modified version of the ANNNI model was
applied to describe  the evolution of the spin and orbital
structure in distorted perovskite manganites. At the same time,
there are a lot of other systems where the competition of nearest-
and next-nearest-neighbor interactions plays an important role,
but where the ANNNI-type approach is hardly applicable to the
actual situation. In this connection, let us note widely discussed
currently magnetic systems with the pyrochlore
structure~\cite{pyrochl} and also spinels. In such structures, a
three-dimensional network of corner-shared cubic cells could be
mapped onto  the square lattice (with holes) having the
nearest-neighbor and diagonal interactions of the same sign. In
principle, such a situation could be treated as a natural
generalization of the ANNNI model to the two-dimensional case.
However, there exists a qualitative difference. Indeed, if here we
take $J_1$ and $J_2$ of the same (antiferromagnetic) sign, we get
the system even more frustrated (we cannot meet the minimum energy
conditions for either of diagonal neighbors, if we meet them for
nearest neighbors and \emph{vice versa}). So, one could expect a
rather interesting and complicated phase diagram  for such a
simply formulated model as an Ising model on a square lattice with
antiferromagnetic nearest and diagonal interactions, where the
spin variable, $s$, has two values, $s = \pm 1$. Despite the
evident simplicity of this model  and its possible importance for
the analysis of different types of superstructures, it received
much less attention than, say, the ANNNI model. At least, we are
unaware of any detailed study of its properties, although the
problem itself was formulated as early as in 1969~\cite{FanPR69}
and the critical properties of such a model were addressed both
numerically and analytically~\cite{LandBindPRB85,GrynbPRB92,
MalakiEPJB06}.

Another interesting example of systems, where the Ising model with
competing antiferromagnetic interactions could be directly
applicable, comes from such currently popular field of research as
arrays of $\pi$-rings~\cite{KusPRL92} and adiabatic quantum
computing~\cite{Farhi}. A single $\pi$-ring is a superconducting
loop consisting of Josephson junctions where at least one of them
is a  $\pi$-junction~\cite{KusPRL92}. Recently such $\pi$-rings
made of a combination of different, high-temperature and
low-temperature, superconducting materials were deposited onto
substrates in the form of one-dimensional and two-dimensional
arrays~\cite{KirtleyNature,KirtleyPRB05}. If there is one or odd
number of $\pi$-junctions in a loop, then the phase shift by $\pi$
in such a junction results in doubly degenerate time-reversed
ground states created in the loop. There arises a persistent
supercurrent circulating in a clockwise or counter-clockwise
direction~\cite{KusPRL92}. Thus, the phase shift by $\pi$ in such
a junction results in the formation of an orbital current or a
magnetic moment at the ring (see \cite{KusPRL92} for details).
Such orbital magnetic moments give rise to a paramagnetic response
of the superconductor, i.e. to the paramagnetic Meissner
effect~\cite{KusPRL92} observed in cuprate
superconductors~\cite{Uppsala,Braunish}. Recently, it was also
shown that the macroscopic ground state of a highly damped
Josephson junction with the PdNi ferromagnetic layer shorted by a
weak link mimics that the $\pi$-ring~\cite{Aprili,Aprili1}. Such a
$\pi$-ring behaves macroscopically as a magnetic nanoparticle with
the quantized flux, the magnetic anisotropy axis being determined
by the junction plane. A chain or a planar array of electrically
isolated $\pi$-rings could be treated as a set of magnetic moments
oriented perpendicular to the plane (Ising spins) and interacting
via magnetic dipole forces (of the antiferromagnetic sign in this
geometry). This dipole-dipole interaction may modify the values of
the orbital magnetic moments and leads to a formation of the
disordered and/or fractal structures in a one-dimensional chain
\cite{Forrester}. Also,  due to this dipole character of the
interaction between the orbital moments, it is necessary to
include into model the next-nearest neighbor interactions in
addition to those between the nearest neighbors.

Due to very interesting properties observed in these systems, they
attract now  a widespread attention, see for example Refs.~
\cite{KirtleyNature,KirtleyPRB05,Forrester2006,Kuerten2005}. To
describe properly the dipole character of the interaction between
the orbital moments we have  included the constant of the
next-nearest neighbor interaction into the model. For the planar
arrays on a square lattice, the next-nearest neighbor interaction
will be a diagonal interaction and as the result the Ising model
with competing interaction arises.

The planar clusters of $\pi$-rings  may be used  in  adiabatic
quantum computations (AQC)~\cite{Farhi} similar to those used with
superconducting flux qubits.   Possible implementation of
adiabatic quantum algorithm with such qubits have been already
achieved  at an effective temperature of 30 mK~\cite{Ilichev2}.
The experimental data are found to be in complete agreement with
quantum mechanical predictions in full parameter space. The idea
of quantum computation by adiabatic evolution is very simple and
based on the original Feynman proposal to use an evolution of
quantum system to find a ground state of certain Hamiltonians like
3D Ising model, which is very difficult to find by other
ways~\cite{Feynman}.

The ground state of a $\pi$-ring cluster depends on the coupling
between the $\pi$-rings. Varying the couplings, one can obtain
different ground states. For the conventional planar array of
$\pi$-rings studied, for example, in Ref.~\cite{KirtleyNature},
the interaction between the individual $\pi$-ring is mainly of the
dipole-dipole character and fixed. However, an introduction of
additional Josephson junction or a current loop located between
the $\pi$-rings or other Josephson loops with persistent current
may change this coupling significantly. For example, an
introduction of an additional Josephson junction between two flux
qubits, each consisting of three Josephson junctions formed a well
controllable coupling between these qubits~\cite{Ilichev}.

The planar clusters of $\pi$-rings may be used for AQC in the
following manner. First, let us notice that such clusters may be
described with the aid of the Ising model with competing
interactions. The value and the type of these interactions depend
on the coupling between $\pi$-rings, which we are going to be able
to control. Next, the problem which is intended to be solved with
the use of the AQC is encoded into the Ising model with the
competing interaction of the finite size lattice associated with a
planar $\pi$-ring cluster. There are many such problems that can
be encoded into the ground state of such Ising model that include,
for example, the distribution of goods and wealth between
customers, the travelling salesman problem, and many others. Each
of them is characterized by its own Ising model with competing
interactions with its own distribution of couplings between the
sites of the square lattice or of the lattice of another type.

The solution of the problem under consideration is related to
finding the ground state of such Ising model of the particular
type. The AQC deals with the adiabatic evolution of the ground
state of the planar system of $\pi$-rings from initial state to
the final state under the slow changing of coupling constants. For
the initial state we choose the conventional checkerboard
antiferromagnetic state where there is only one type of coupling
constant - the nearest-neighbor antiferromagnetic one. The final
state is associated with a specified distribution of the coupling
constants related to the given problem. The reading out the
distribution of the orientations of the orbital moments of the
$\pi$-rings in the final state will produce a solution of the
given problem.

Thus, the implementation of AQC with the use of the planar
clusters of $\pi$-rings is straightforward. The matter is only to
find and to determine the possible ground states arising in the
Ising model with competing interactions at the different values of
the coupling constants.

The plan of the present paper is the following. First, we consider
small planar clusters of $\pi$-rings, which constitutes portions
of the square lattice. These clusters consist of four, five, 16,
and 25 $\pi$-rings. If a single $\pi$-ring is associated with a
site with the Ising spin, then, for example, one of the considered
clusters, the 5-site cluster, consists of  a central site with its
four neighbors. Such a 5-site cluster or a plaquette correctly
reproduces the symmetry of the square two-dimensional lattice and
large clusters.  The exact solution of such a plaquette is easily
found, and the physical mechanisms underlying the the main
features of its behavior are discussed. Then, based on a
mean-field type approach, we discuss possible spin configurations
and the types of topological defects characteristic of the
infinite square lattice of Ising spins. At the end of the paper,
we compare the obtained results with the Monte Carlo simulations
and with the experimental data for the arrays of $\pi$-rings. We
also briefly discuss the possible applications of the systems
under study to the adiabatic quantum computation.

\section{Model}

As it was discussed in Introduction, we start from the
two-dimensional Ising model with antiferromagnetic
nearest-neighbor and diagonal interactions. The Hamiltonian for
such a model can be written as

\begin{equation}\label{eq:1} H = J\sum\limits_{\langle i,
j\rangle_{nn}}s_is_j +J'\sum\limits_{\langle i,
j\rangle_{dn}}s_is_j \end{equation}

\begin{figure}[htb] \centering
\includegraphics[width=0.4\textwidth]{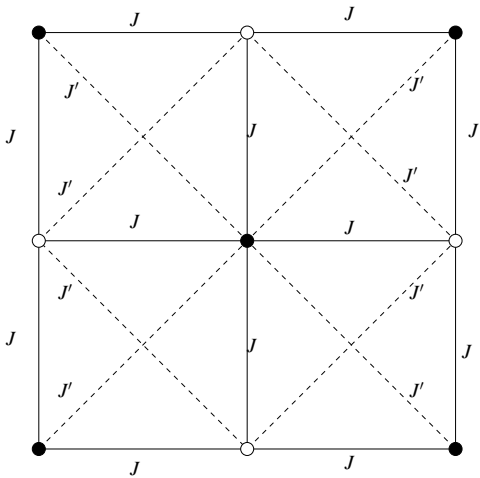} \caption{Ising
model with nearest neighbor $J$ and diagonal $J'$ interactions for
the square lattice. Filled and open circles mean $s = +1$ and $s =
-1$, respectively. Here, the usual two-sublattice arrangement of
spins is shown.} \label{Fig1} \end{figure}

Here $J, J' > 0$, $s$ is a two-value Ising variable $s = \pm 1$,
$\langle i, j\rangle_{dn}$ and $\langle i, j\rangle_{dn}$ denote
the summation over sites $i$ and $j$  being respectively nearest
neighbors ($nn$) and diagonal neighbors ($dn$). The geometry of
the model is schematically illustrated in Fig. \ref{Fig1}.

Such a model correctly describes a planar array of $\pi$-rings
deposited onto an insulating substrate. At each $\pi$-ring, there
arises a magnetic orbital moment - an Ising spin. Such moments are
oriented perpendicular to the plane and therefore they interact
with each other via the long-range antiferromagnetic (dipole)
interaction.  Since such interactions decrease with a distance
between moments as $~1/r^3$, i.e. very fast, we need to consider
only interactions between the nearest and next-nearest neighboring
spins, i.e. in the square lattice, there will be two constants of
antiferromagnetic interaction, $J$ and $J'$. Of course, due to the
dipolar character of interaction the values of $J$ and $J'$ for
arrays of $\pi$-rings are related as $J'=J/2^{3/2}$, but we
consider a more general case, when their values are arbitrary.

\begin{figure}[htb] \centering
\includegraphics[width=0.4\textwidth]{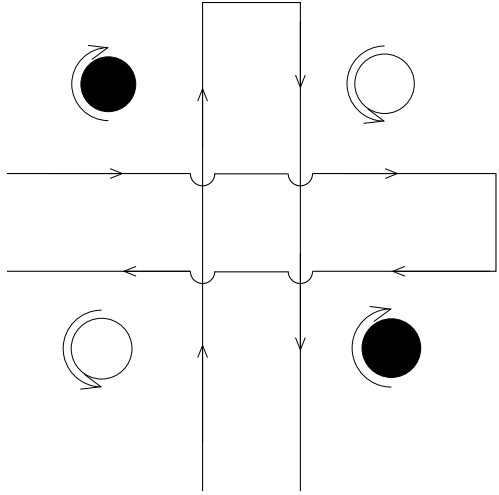}
\caption{The smallest square cluster of $\pi$-rings, $2\times2$.
Here the value and the ratio of nearest-neighbor $J$ and diagonal
$J'$ interactions can be controlled by the external bias current
$I$. The directions of currents in the $\pi$-rings and in control
loops are indicated by arrows. Filled and open circles denote
clockwise and counter-clockwise directed orbital currents
associated with the orbital moments equal to $s = +1$ and $s =
-1$, respectively.} \label{Fig2} \end{figure}

For the planar systems of $\pi$-rings, the values of interactions
$J$ and $J'$ and their ratio can be externally controlled.  One of
the many possible way to control these interactions in the planar
$\pi$-ring arrays is to introduce additional currents or current
loops between the $\pi$-rings (see, Fig.\ref{Fig2}). In this
figure, we consider a simplest square plaquette of $\pi$-rings. If
the currents are flowing along the pairs of the straight lines in
opposite directions, the magnetic field induced by these currents
will influence the interaction between the $\pi$-rings and the
ratio of the values $J/J'$ will be changed. In a similar way, a
controllable coupling of two loops, each consisting of the
three-junction flux qubits, has been realized by inserting an
additional current loop between them~\cite{Ilichev}. The
$\pi$-ring is usually considered a primary candidate for the flux
qubit due to a long decoherence time and low noises.  The detailed
measurements presented in Ref.~\cite{Ilichev} illustrate the
flexibility of this two-loop tunable coupling that may be easily
changed in a very broad range. The controllable adiabatic
evolution of the multi $\pi$-ring systems, which we consider here,
can be used for adiabatic quantum computation (AQC), see, for
example, Ref.~\cite{Ilichev1}, where the first experimental
realization of AQC with the use of the controllable coupling has
been realized. Bearing these experiments in mind, we consider
small planar clusters of $\pi$-rings and analyze the diagrams of
possible states for these small clusters at different values of
the ratio $J/J'$, which can arise via a controllable coupling.

\section{Thermal properties of small square plaquettes}

\subsection{A $2\times2$ plaquette}

The simplest cluster of $\pi$-rings, which may be easily built up
experimentally is the square $2\times2$ plaquette having four
$\pi$-rings. The similar plaquette of $2\times 2$ superconducting
loops  has been build up and tested  experimentally in the
Ref.~\cite{Ilichev3}. There was the first experimental
demonstration of the working AQC device having the four qubits
with a mixed couplings.  The working circuit demonstrated in the
Ref.~\cite{Ilichev3} consisted of four three-junction loops --
four flux qubits, with simultaneous ferro- and antiferromagnetic
coupling implemented using shared Josephson junctions.  The result
obtained in the Ref.~\cite{Ilichev3} gave a confidence that the
similar working circuit based on four $\pi$-rings can be also
implemented.

Such a $\pi$-ring cluster has $2^4=16$ distinct configurations
that we label $0 \ldots 15$. Converting these state labels to
binary we have

\begin{table}[htbp] \begin{center} \begin{tabular}{|l|c|c||l|c|c|}
\hline
    Configuration & State & State & Configuration & State & State \\ \hline
    0 & 0000 & +\ +\ +\ + & 8 & 1000 & -\ +\ +\ + \\
    1 & 0001 & +\ +\ +\ - & 9 & 1001 & -\ +\ +\ - \\
    2 & 0010 & +\ +\ -\ + & 10 & 1010 & -\ +\ -\ + \\
    3 & 0011 & +\ +\ -\ - & 11 & 1011 & -\ +\ -\ - \\
    4 & 0100 & +\ -\ +\ + & 12 & 1100 & -\ -\ +\ + \\
    5 & 0101 & +\ -\ +\ - & 13 & 1101 & -\ -\ +\ - \\
    6 & 0110 & +\ -\ -\ + & 14 & 1110 & -\ -\ -\ + \\
    7 & 0111 & +\ -\ -\ - & 15 & 1111 & -\ -\ -\ - \\ \hline
\end{tabular} \caption[Possible configurations of a $2\times2$
lattice]{Possible configurations of a $2\times2$ lattice
illustrated by converting the binary form of the spin
configuration into spin states } \end{center} \end{table}

\begin{figure}[htb] \centering
\includegraphics[width=0.7\textwidth]{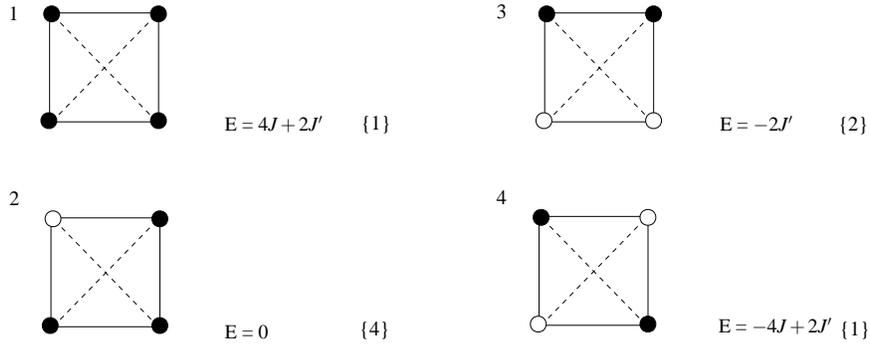}
\caption{Possible nonequivalent spin configurations of a
$2\times2$ plaquette and the corresponding energies; the degree of
degeneracy of each energy value is shown in curly brackets. The
other 8 configurations are obtained by the reversal of all spins
in the plaquette.} \label{Fig3} \end{figure}

\begin{figure}[htb] \centering
\includegraphics[width=0.7\textwidth]{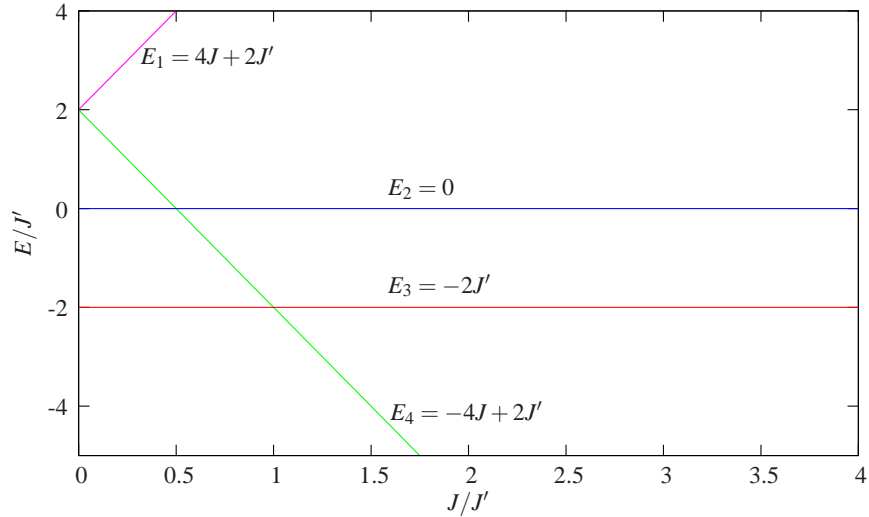}
\caption{Energies of different states of a $2\times2$ plaquette
versus parameter $J/J'$.} \label{Fig4} \end{figure}

If there are only two types of coupling, $J$ and $J'$, between all
$\pi$-rings these states are associated with 4 different energies:
1) for the ferromagnetic state, $E_{++++}=4 J +2 J'$; 2) for the
one spin flop, $E_{+++-}=0$; 3) for the stripe order,
$E_{++--}=-2J'$ and 4) for the antiferromagnetic order,
$E_{+-+-}=-4J+2J'$. The corresponding spin configurations and the
dependence of their energies on $J/J'$ are shown in Figs.
\ref{Fig3} and \ref{Fig4}. At zero temperature, either the stripe
order arising at $J/J'<1$ or the antiferromagnetic or the
checkerboard order, arising at $J/J'>1$ will dominate.  At zero or
very low temperatures at slow changes of this ratio, the ground
state adiabatically evolves from the checkerboard to the stripe
order. The effective transition happens when $J=J'$.

\begin{figure}[htb] \centering
\includegraphics[width=0.6\textwidth]{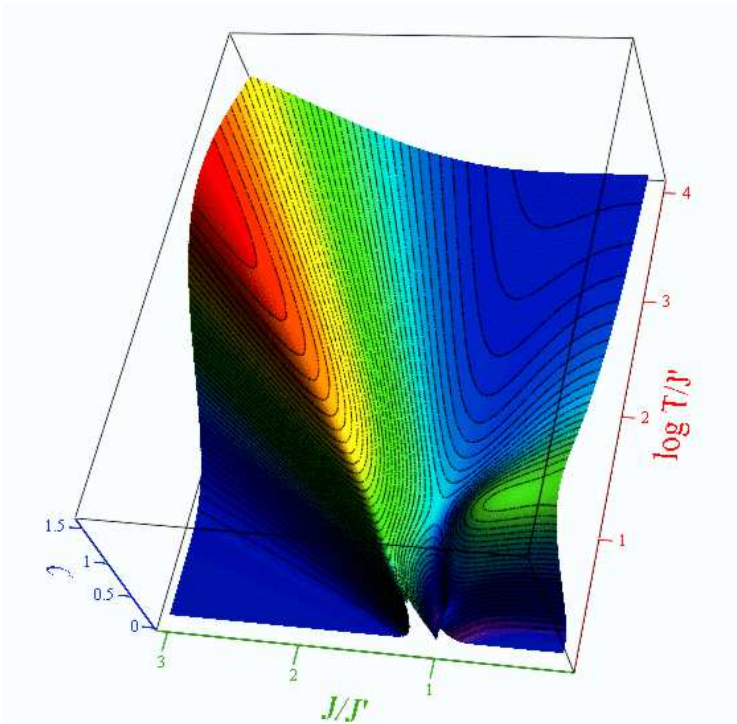}
\caption{ (color online) A three-dimensional plot of heat capacity
$C$ versus $T/J'$ and $J/J'$ for a $2\times2$ plaquette.}
\label{Fig5}
\end{figure}

For a description of the evolution at any fixed temperature, we
have to determine the partition function and the free energy of
the system. We can populate our $2\times2$ lattice with each of
these enumerated states and sum up each contribution to get the
partition function. Next, we find the free energy and the specific
heat. Its maxima at the different ratio $J/J'$ and the temperature
$T$ give a phase diagram. This diagram indicates that such a toy
adiabatic quantum computer operates well in the broad range of
temperatures performing the adiabatic computation from the
checkerboard to the stripe order, where the transition arising at
the value $J=J'$ is slightly smeared (see Fig.~\ref{Fig5}).

In a more general situation at the different coupling between
different $\pi$-rings, any state enumerated in Table 1 may become
a ground state that can be associated with some realistic problem
needed to be solved with the AQC.

\subsection{A 5-site plaquette: ground state and possible
configurations}

\begin{figure} \centering
\includegraphics[width=0.3\textwidth]{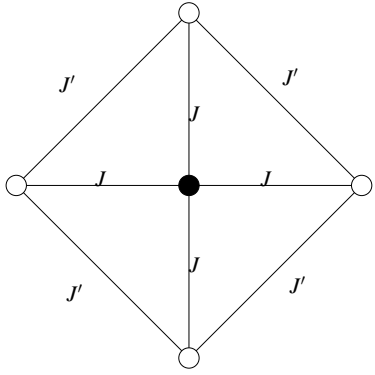}
\caption{Square plaquette of the minimum size (with the equal
number of of nearest-neighbor and diagonal bonds).}
\label{Fig6}
\end{figure}

Of course, to go beyond such a toy computer we have to consider
larger systems with large variety in the distribution of the
coupling constants between the $\pi$-rings. We found that the
behavior obtained for larger clusters is very different from one
obtained above. The next system which, nonetheless, may already
have nontrivial generic features is a cluster (plaquette)
consisting of five $\pi$-rings. Its central cite has four
neighbors.  For the further analysis, we choose this plaquette of
minimum size, because it retains the symmetry and the
characteristic geometry of the lattice as a whole (the number of
nearest-neighbor and diagonal bonds are equal, see
Fig.~\ref{Fig6}). The properties of such plaquette consisting of 5
$\pi$-rings  may be generic to develop a scalable architecture for
AQC.

\begin{figure} \centering
\includegraphics[width=0.8\textwidth]{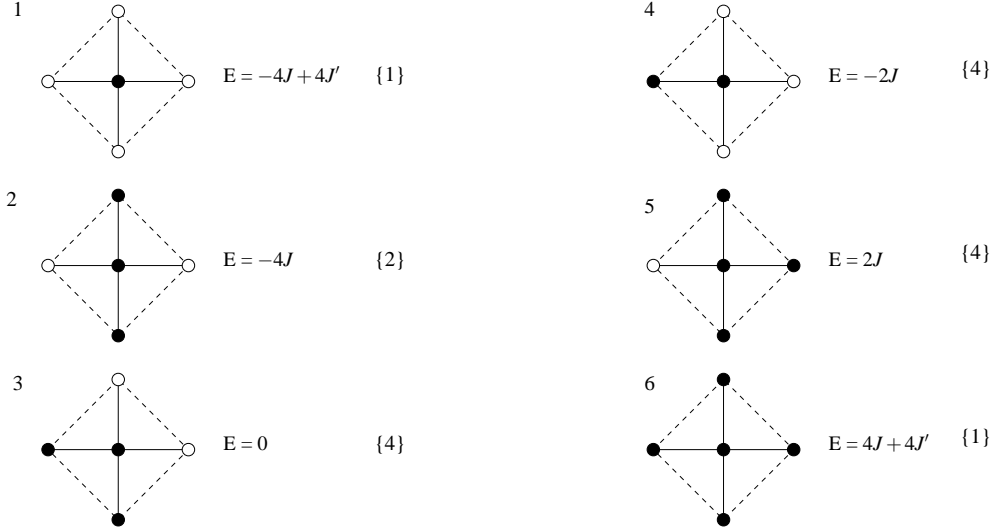}
\caption{Possible spin configurations of a 5-site plaquette and
the corresponding energies; the degree of degeneracy of each
energy value is shown in curly brackets. Only 16 configurations
with the filled circle in the center are shown; there are also 16
similar configurations with the open circle in the center of the
plaquette.}
\label{Fig7}
\end{figure}

\begin{figure}\centering
\includegraphics[width=0.8\textwidth]{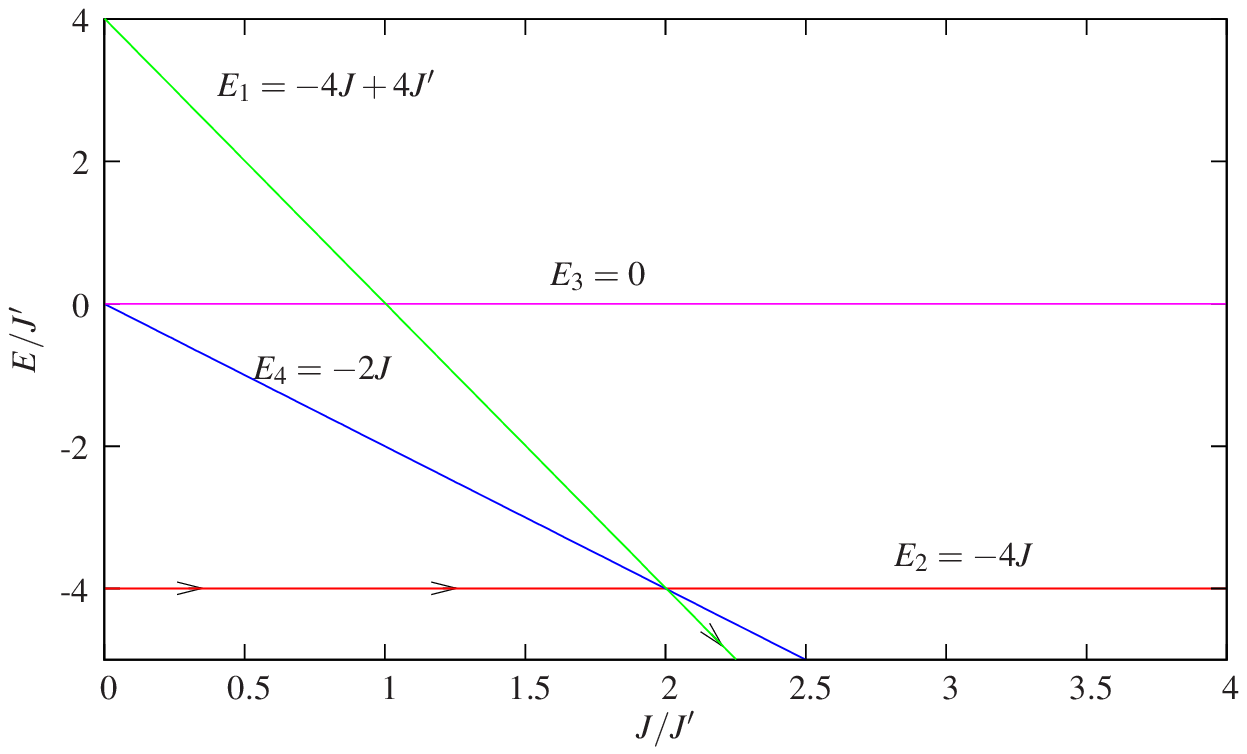}
\caption{Energies of different states of a 5-site plaquette versus
parameter $J/J'$. The solid line with arrows illustrates the
variation of parameters corresponding to the adiabatic
computations.} \label{Fig8}
\end{figure}

The possible configurations of this 5-site plaquette are easily
listed (we have $2^5 = 32$ configurations with different values of
energy), see Fig. \ref{Fig7}. In Fig. \ref{Fig8}, we plot the
dependence of energies for the lowest energy states (states 1, 2,
3, and 4) on the $J/J'$ ratio.

One can see that state 1 corresponding to the usual two-sublattice
antiferromagnetism (the checkerboard arrangement of filled and
open circles) is favorable at $J/J' > 2$, whereas state 2
corresponding to the alternation of chains formed by filled and
open circles becomes favorable at rather large magnitude of the
diagonal interaction, $J/J' < 2$. Note that for state 4, plot
$E_4(J/J')$ also passes through the crossing point of $E_1(J/J')$
and $E_2(J/J')$. State 4 could be treated  as a defect in regular
lattices corresponding to states 1 and 2. So, the crossing of
plots at $J/J' = 2$ gives a clear indication of a possible
formation of some more complicated phases near this point. Indeed,
an almost zero barrier to the formation of defects, like domain
boundaries and dislocations, is usually a good signature of the
situation when some kind of superstructure could be favorable. For
the case of AQC, the computation process may start from the
$\pi$-ring cluster, where the value of $J/J'>2\sqrt{2}$. There,
the ground state must correspond to the plain checkerboard
antiferromagnetic order. If we slowly decrease the $J/J'$ ratio
the system will inevitably evolve to the stripe structure. This
demonstrates the proper operation of the AQC having 5 qubits.
However, at nonzero temperatures the situation is much more subtle
than that.

\subsection{A 5-site plaquette: partition function, free energy,
and peaks in specific heat}

Now, let us discuss the behavior of the 5-site plaquette at finite
temperatures. Knowing the set of energy levels  and their
degeneracy (see Fig. \ref{Fig7}), we can easily write  the
partition function in the form (the Boltzmann constant is taken to
be equal to one)

\begin{eqnarray}\label{eq:2}
Z = 2\left[  e^{\frac{4J-4J'}{T}} +
2e^{\frac{4J'}{T}} + 4 +
4e^{\frac{2J}{T}} + 4e^{\frac{-2J}{T}} +  e^{\frac{-4J-4J'}{T}}\right] = \nonumber\\
= 8 \left[ e^{\frac{-4J'}{T}}\cosh \frac{2J}{T} + 2\cosh
\frac{2J}{T} + 2\sinh \frac{4J'}{T} + 1 \right].
\end{eqnarray}

Free energy $F$ and heat capacity $C$ for the plaquette are given
by standard thermodynamic formulas $F = -T\ln Z$ and $C =
-T\frac{\partial^2 F}{\partial T^2}$. Then, the peaks in the
$C(T)$ curves at different values of $J/J'$ can be treated as
manifestations of phase transitions, which are going to occur for
the infinite lattice. From this viewpoint, let us analyze the
low-temperature behavior of the heat capacity.

From the usual expression for the heat capacity
\begin{equation}\label{eq:3}
 C(T) = -T\frac{\partial^2 F}{\partial T^2} =
\frac{2T}{Z}\frac{\partial Z}{\partial T} -
\frac{T^2}{Z^2}\left(\frac{\partial Z}{\partial T}\right)^2 +
\frac{T^2}{Z}\frac{\partial^2 Z}{\partial T^2},
\end{equation}
we can see that the structure of the $C(T)$ function is such that
it is possible to get rid of the growing exponentials in the
partition function, $Z$, by dividing both numerator and
denominator in $C(T)$ by the largest exponential. Then, we can
work with exponentials having negative arguments. Indeed, let
\begin{equation}\label{eq:4}
 Z = A'e^{\frac{A}{T}}(1 + Be^{-\frac{D}{T}} + \ldots).
\end{equation}
From  Eq. (\ref{eq:3}), it obvious that $C(T)$ does not depend
either on $A$ or on $A'$. In the low-temperature limit, taking
into account only the terms proportional to $e^{-\frac{D}{T}}$, we
find
\begin{equation}\label{eq:5}
 C(T)\approx \frac{BD^2}{T^2}e^{-\frac{D}{T}}.
\end{equation}
Note that the terms $\sim e^{-\frac{D}{T}}/T$ are
cancelled. Using expression (\ref{eq:5}) and condition $dC/dT =
0$, we find temperature $T_c$ corresponding to the position of the
heat capacity peak (in the limit $T\rightarrow 0$)
\begin{equation}\label{eq:6}
 T_c = \frac{D}{2}.
\end{equation}

Let us introduce for convenience the following dimensionless
variables
 \begin{equation}\label{eq:7}
 f = \frac{F}{J'}, \quad t = \frac{T}{J'}, \quad \alpha = \frac{J}{J'}.
\end{equation}

In this notation,  Eq. (\ref{eq:2}) for the partition function can
be rewritten in the following way

\begin{equation}\label{eq:8} Z = 2\left[ e^{\frac{4\alpha - 4}{t}}
+ 2e^{\frac{4}{t}} + 4
 + 4e^{\frac{2\alpha}{t}} + 4e^{\frac{-2\alpha}{t}}
 + e^{\frac{-4-4\alpha}{t}}\right].
\end{equation}

As it was mentioned above, to analyze the low-temperature limit,
it is reasonable to retain only growing exponentials in Eqs.
(\ref{eq:2}) or (\ref{eq:8}). Thus, we get
\begin{equation}\label{eq:9}
Z \approx  2\left[ e^{\frac{4\alpha - 4}{t}} + 2e^{\frac{4}{t}}
 + 4e^{\frac{2\alpha}{t}}\right].
\end{equation}

At $\alpha = \frac{J}{J'} > 2$, we can write the partition
function in the form similar to (\ref{eq:4})
\begin{equation}\label{eq:10}
 Z \approx 2e^{\frac{4J - 4J'}{T}}(1 + 4e^{-\frac{D_1}{T}} + \ldots),
\end{equation} where $D_1 = 2J - 4J'$.

In the case $\alpha = \frac{J}{J'} < 2$, we find
\begin{equation}\label{eq:11}
 Z \approx 4e^{\frac{4J'}{T}}(1 + 4e^{-\frac{D_2}{T}} + \ldots),
\end{equation} where $D_2 = 4J' - 2J$.

As a result, using (\ref{eq:6}), we find the following positions
of the heat capacity peaks in the low-temperature limit
\begin{equation}\label{eq:12} \alpha = 2 + t, \quad \alpha > 2.
\end{equation} \begin{equation}\label{eq:13} \alpha = 2  - t,
\quad \alpha < 2. \end{equation}

\begin{figure}\centering
\includegraphics[width=0.8\textwidth]{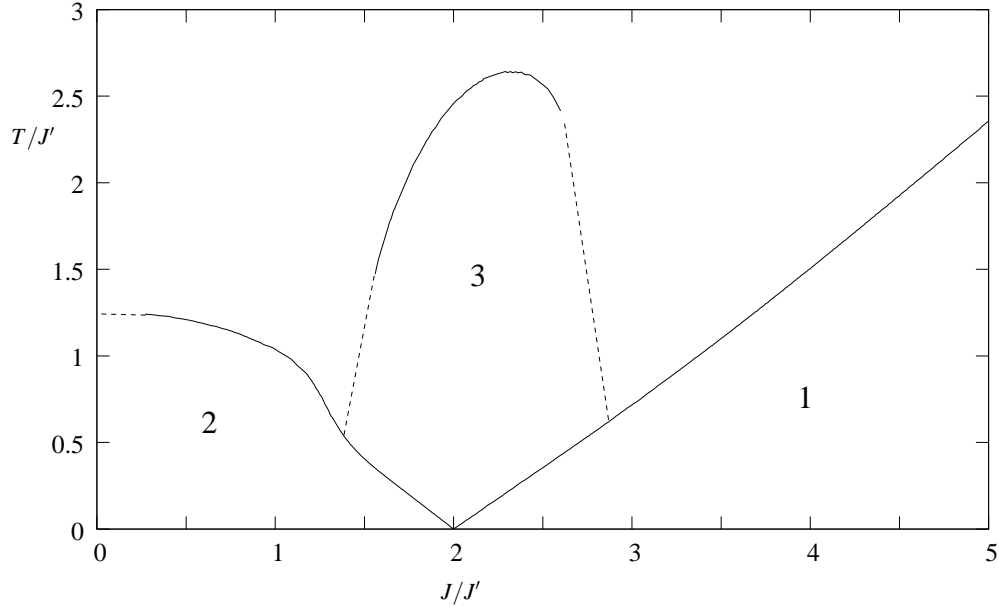}
\caption{``Phase diagram''(positions of the heat capacity peaks)
for a 5-site plaquette in the $t-\alpha$ plane. Region 1
corresponds to state 1 of the plaquette (see Fig.~\ref{Fig7}),
region 2 could be related to state 2 of the plaquette. In region
3, one could expect for the corresponding infinite square lattice
multiple phases (similar to those in the ANNNI model) or a phase
with a complicated spin ordering.}
\label{Fig9}
\end{figure}

\begin{figure} \centering
\includegraphics[width=0.6\textwidth]{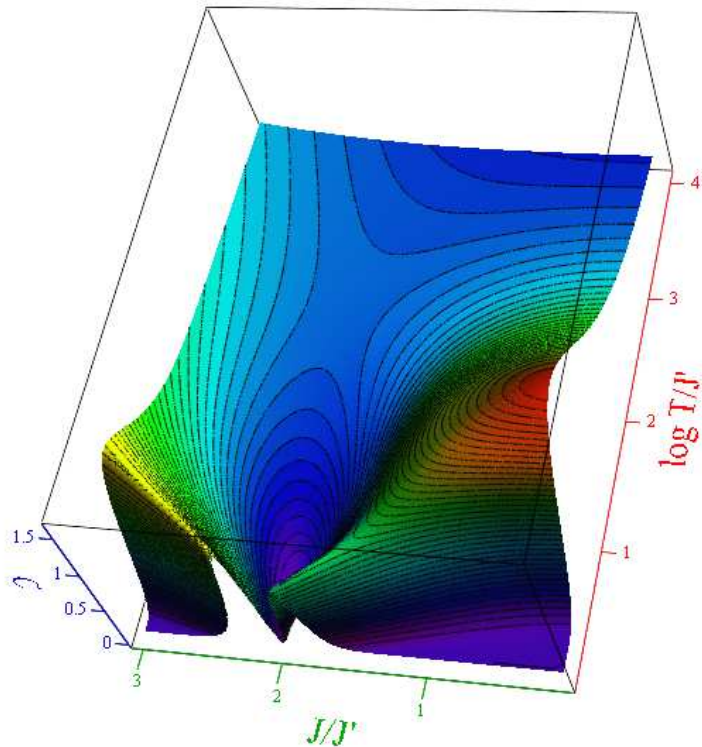}
\caption{A three-dimensional plot of heat capacity $C$ versus $t =
T/J'$ and $\alpha = J/J'$, which corresponds to the phase diagram
shown in Fig. \ref{Fig9}.} \label{Fig10}
\end{figure}

Plotting the positions of heat capacity maxima in the $t-\alpha$
plane yields a kind of phase diagram  for a small plaquette.
Indeed, the positions of $C(t, \alpha)$ peaks could be related to
the actual phase  transitions occurring in infinite systems. Such
a phase diagram for the plaquette under study is shown in Fig.
\ref{Fig9}. Note that at $\alpha < 2$, the lowest energy
corresponds to plaquette 2 in Fig. \ref{Fig7} (chains of spins
with the same direction), whereas at $\alpha > 2$, plaquette 1
(two-sublattice checkerboard antiferromagnetism) is the most
favorable. However, at non-zero temperatures, there is no boundary
between these two phases. Instead, we have a whole quadrant in the
$t-\alpha$ plane (with the vertex at point (0, 2)), where one
could expect for the infinite lattice numerous phases,
intermediate between the two-sublattice and chain-like structures.
The situation is very much alike that of the ANNNI model
\cite{SelkePRep}, where the ``devil's staircase'' of different
phases arises at finite temperatures between two main phases
existing at zero temperature. Region 3 within this quadrant is
bounded by a curve corresponding to the positions of an additional
(high-temperature) peak in the temperature dependence of the heat
capacity. So, this is another indication that for larger clusters,
and in the infinite system, one should expect in this range a
phase with a certain kind of ordering, maybe quite unusual, rather
than simply a disordered paramagnetic phase. For illustration, we
present here also a three-dimensional plot of heat capacity $C(t,
\alpha)$ (see Fig.~\ref{Fig10}) corresponding to the phase diagram
shown in Fig.~\ref{Fig9}.

To summarize, when the ratio $J/J'$ is in the vicinity of the
value 2 for the plaquette cluster of five $\pi$-rings, the
antiferromagnetic, checkerboard ground state (1) having zero
entropy (nondegenerate) is very difficult to reach. So, the system
is practically always disordered due to a proliferation of defects
(4), see, Figs. \ref{Fig3} and \ref{Fig8}. This
diagram also indicates that some problems arise even for the toy
5-$\pi$-ring adiabatic quantum computer. Although it operates well
at zero temperature, in very broad range of low temperatures
performing adiabatic computations such as the adiabatic evolution
from the checkerboard to the stripe order may have an obstacle in
the form of the proliferation of topological defects inherently
presented in the phase (3). Now, we would like to show that these
difficulties with AQC even are enhanced in the case of larger
clusters. Also, the broad peak in specific heat indicates that a
true phase transition may arise for a large system of $\pi$-rings.

\section{Exact solutions for larger plaquettes}

In the previous Section, we found all possible energy states for
the 5-site plaquette, see Fig. \ref{Fig7}. For a small portion of
the square lattice (less than about $4\times4$ sites), it is
possible to list its different spin configurations and calculate
the macroscopic quantities exactly. For a $3\times3$ site lattice
we find that there are 11 different energies associated with $2^9$
different states and can rather easily reproduce the analysis of
the minimum of energy but due to its small size this is of limited
value. For a $5\times5$-site square, there are already 161
different energy levels and thus the approach used in the previous
Section seems to be impractical.

\begin{figure} \centering
\includegraphics[width=0.7\textwidth] {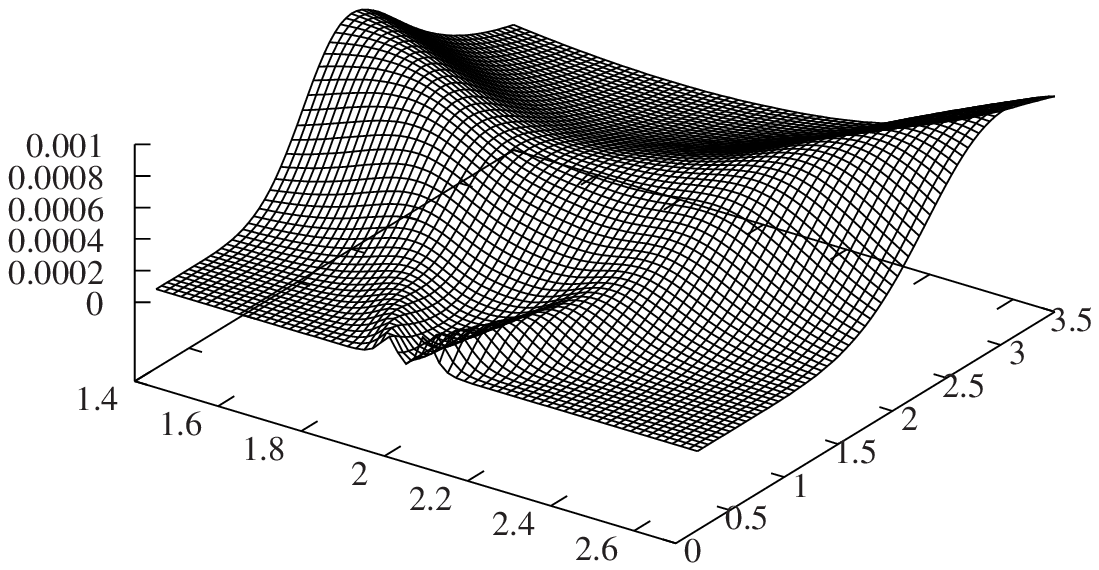}
\caption{A three-dimensional plot of heat capacity $C$ versus $t =
T/J'$ and $\alpha = J/J'$ for a $4\times4$ square plaquette.}
\label{Fig11}
\end{figure}

\begin{figure} \centering
\includegraphics[width=0.7\textwidth]{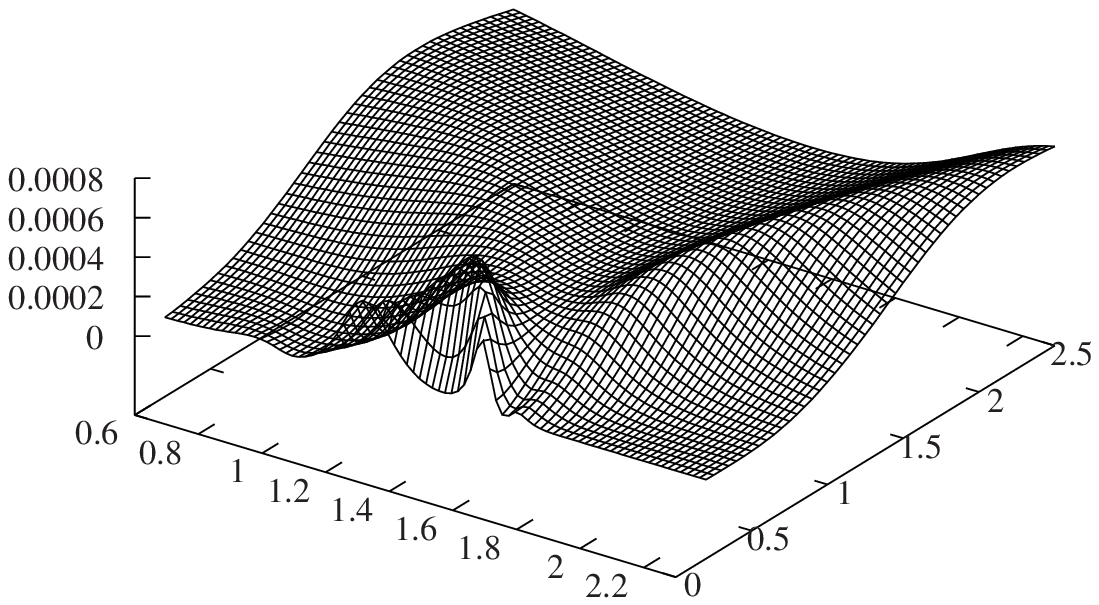}
\caption{A three-dimensional plot of heat capacity $C$ versus $t =
T/J'$ and $\alpha = J/J'$ for a $5\times5$ square plaquette.}
\label{Fig12}
\end{figure}

A $5\times5$-site lattice has $2^{25}\approx3\times10^7$ distinct
configurations that can be enumerated  by converting each state
($0 \ldots 2^N-1$) into a binary number where the 0 or 1 digit of
the binary form corresponds to an up or down spin in our
Hamiltonian (similarly to the notation introduced in Table 1 for
the $2\times2$ plaquette). With the use of such enumeration, we
calculate the partition function function of the lattice and thus
deduce the macroscopic quantities of interest by numerically
differentiating the partition function using the usual finite
difference schemes. As a result, for larger square plaquettes
(say, with $4\times4$ and $5\times5$ sites), we can find their
heat capacity and draw 3D plots similar to that shown in
Fig.~\ref{Fig10}, which illustrate the phase diagram of these
plaquettes. The corresponding plots for $4\times4$- and
$5\times5$-site squares are shown in Figs.~\ref{Fig11} and
\ref{Fig12}. The qualitative features of these plots reproduce the
situation characteristic of the 5-site plaquette discussed in the
previous Section. Note, however, that cutting from the infinite
square lattice finite $N\times N$-site pieces (with the cuts
parallel to $x$ and $y$ axes), we get square plaquettes with
unequal number of $J$ and $J'$ bonds. The corresponding ``surface
contribution" can affect the specific form of the phase diagram,
especially, in the regime of the strong degeneracy, when $J/J'=2$.

\section{Square infinite lattice: semiqualitative analysis
 and energetics of topological defects}

\begin{figure}\centering
\includegraphics[width=0.8\textwidth]{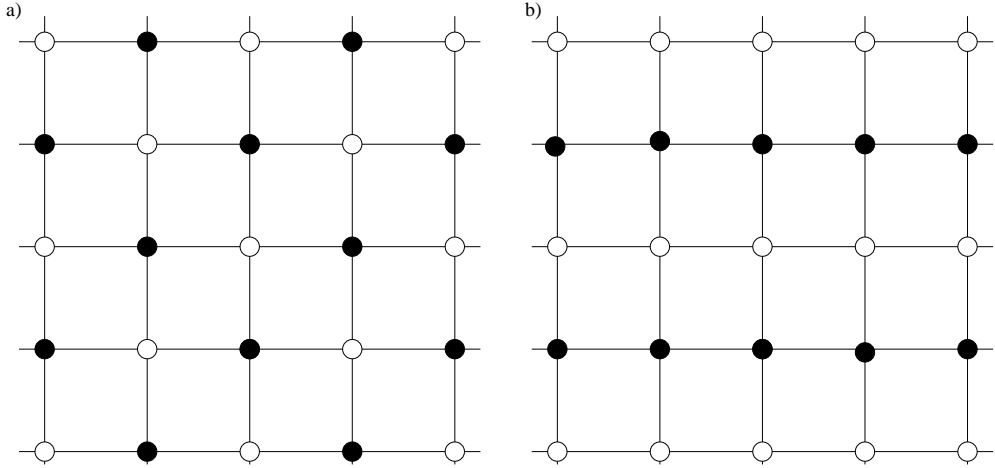}
\caption{Two types of spin ordering corresponding to the lowest
energy  for the square lattice with the the antiferromagnetic
nearest-neighbor $J$ and diagonal $J'$ interactions: (\emph{a})
two-sublattice classical antiferromagnetism (diagonal stripes)
arising when $J/J' \gg 1$ and (\emph{b}) chain-like
antiferromagnetic ordering (horizontal stripes) arising when
$J/J'<<1$. The rotation of configuration (\emph{b}) by 90$^\circ$
gives vertical stripes with the same energy.}\label{Fig13}
\end{figure}

Let us discuss the possible spin configurations corresponding to
Hamiltonian (\ref{eq:1}) for the infinite square lattice. Two
configurations, which may have the lowest energy per site at large
and small ratio of coupling constants $J/J'$  are shown in Fig.
\ref{Fig13}.

For diagonal stripes, we gain energy at nearest-neighbor
interactions and loose energy at diagonal bonds, hence this
configuration is the most favorable at $J/J' \gg 1$. For the
horizontal (vertical) stripes, the main energy gain  comes from
the diagonal neighbors, and it becomes favorable at large values
of $J'$ ($J/J' \ll 1$). The corresponding energies (per site) are

\begin{equation}\label{eq:14} E_{ds}= -2J+2J' \quad \text{for
diagonal stripes}, \end{equation}

\begin{equation}\label{eq:15} E_{hs}= -2J' \quad \text{for
horizontal (vertical) stripes}. \end{equation}

The transition between diagonal and horizontal stripe states
($E_{ds}=E_{hs}$) occurs at

\begin{equation}\label{eq:16} J/J' =2. \end{equation}

Let us now discuss the possible types of defects and phase
boundaries, which could arise in our model.

\begin{figure}\centering
\includegraphics[width=0.8\textwidth]{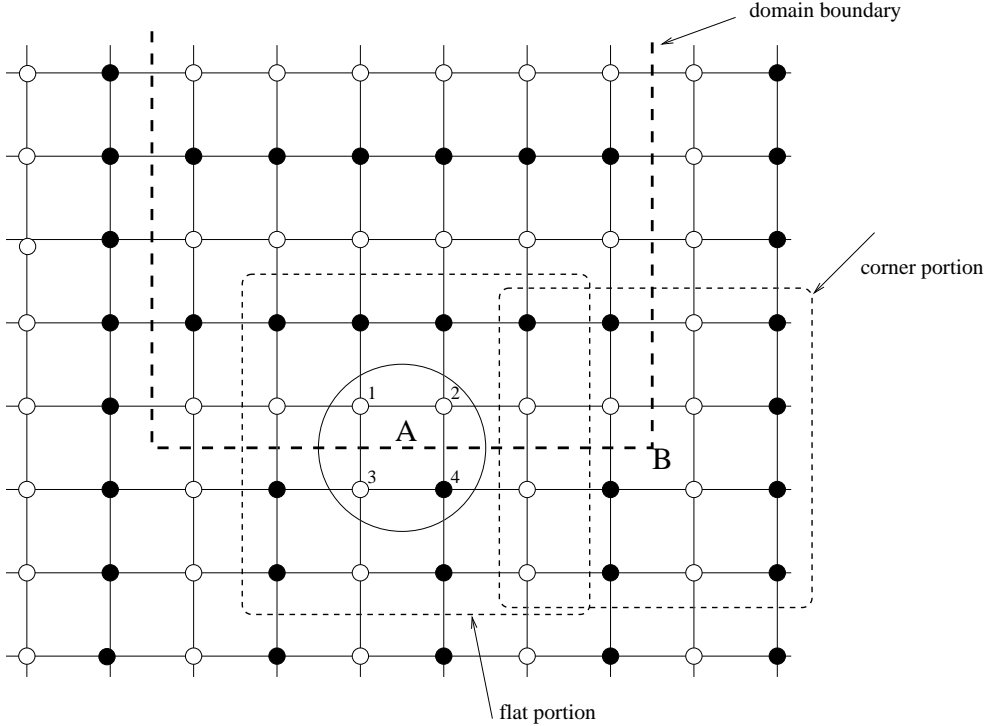}
\caption{Domain boundary between horizontal and vertical stripes
(horizontal or vertical boundary).}
\label{Fig14}
\end{figure}

First, let us analyze the state with horizontal or vertical
stripes. The most natural defect for this state is just the
boundary between the domains corresponding to vertical and
horizontal stripes (see Fig. \ref{Fig14}).

Let us now calculate the energy (per site) for the flat portion of
this domain boundary. We take a 4-site plaquette around point A in
Fig. \ref{Fig14} and calculate  energies of sites numbered  from 1
to 4. Thus, we have

\begin{eqnarray}\label{eq:17}
E_1 = \frac{1}{2}\left[-4J' + 3J - J \right] = -2J' + J,\nonumber\\
E_2 = \frac{1}{2}\left[-2J' + 2J' - 2J + 2J\right] = 0,\nonumber\\
E_3 = \frac{1}{2}\left[-2J' + 2J' + 2J - 2J\right] = 0,\nonumber\\
E_4 = \frac{1}{2}\left[-4J' + J - 3J \right] = -2J' - J,\nonumber\\
E_{fp}= \frac{1}{4}\left[E_1 + E_2 + E_3 +E_4 \right] = -J'.
\end{eqnarray}

The same calculation can be performed for the corner portion of
this domain boundary (see four sites surrounding point B in Fig.
\ref{Fig8}). After the procedure similar to that presented by Eq.
(\ref{eq:17}), we find

\begin{equation}\label{eq:18} E_{cp}= -J'. \end{equation}

So, we see that $E_{cp}=E_{fp}$, that is, kinks at the domain
boundary  do not cost additional energy. This means that at finite
temperatures, the formation of polygonal domain boundaries could
be favorable from the viewpoint of the configurational entropy.

\begin{figure} \centering
\includegraphics[width=0.8\textwidth]{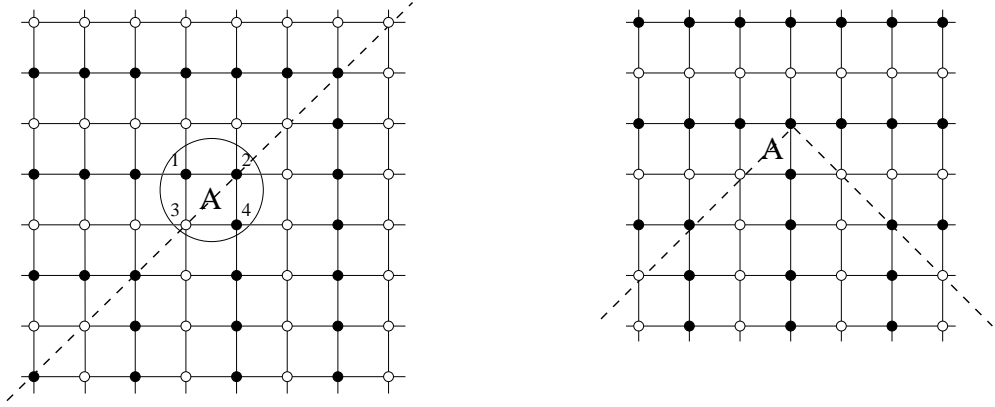}
\caption{Diagonal domain boundary for the phase with horizontal
and vertical stripes (left panel) and the corner portion for this
type of domain boundary (right panel).}
\label{Fig15}
\end{figure}

In addition to the domain boundaries along the vertical and
horizontal axes, it is possible to consider also a diagonal domain
boundary illustrated in Fig. \ref{Fig15}.

Again, calculating the energies of four sites surrounding point A
in the left panel of Fig. \ref{Fig15}, we find

\begin{eqnarray}\label{eq:19}
E_1 = \frac{1}{2}\left[-2J + 2J - 3J' + J'\right] = -J',\nonumber\\
E_2 = \frac{1}{2}\left[-2J + 2J - 4J' \right] = -2J',\nonumber\\
E_3 = \frac{1}{2}\left[-2J + 2J - 4J' \right] = -2J',\nonumber\\
E_4 = \frac{1}{2}\left[-2J + 2J - 3J' + J' \right] = -J',\nonumber\\
E_{fp}= \frac{1}{4}\left[E_1 + E_2 + E_3 +E_4 \right] =
-\frac{3}{2}J'. \end{eqnarray}

So, we see that the diagonal domain boundary  has even lower
energy than the vertical and horizontal boundaries. The corner
portion of this kind of domain boundary is illustrated in the
right panel of Fig.~\ref{Fig15}. The energy per site for this
corner portion can be found by considering  the neighbors of site
A in the right panel of Fig.~\ref{Fig15}. Thus, we have

\begin{equation}\label{eq:20} E_{cp-d} = -2J' + J. \end{equation}

At large $J'$, this energy can be even smaller than that given by
Eq. \ref{eq:19}.

\begin{figure}\centering
\includegraphics[width=0.8\textwidth]{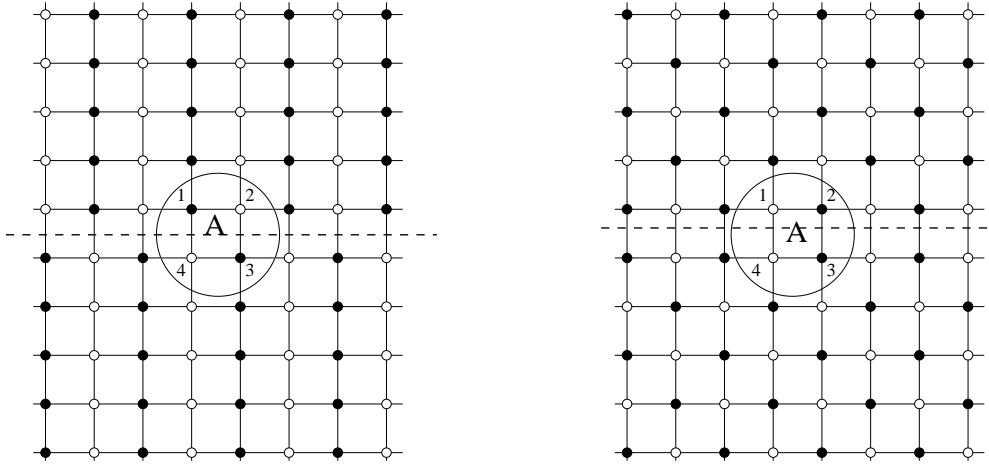}
\caption{A dislocation(shift by one lattice constant) in the
phases with horizontal and vertical stripes (left panel) and with
diagonal stripes (right panel).}
\label{Fig16}
\end{figure}

In addition to the domain boundaries, there are also dislocations
of the type shown in Fig.~\ref{Fig16}.

Calculating the energy of four spins around point A in the left
panel of Fig.~\ref{Fig16}, we find that the energy of such a
dislocation per site is

\begin{equation}\label{eq:21}
 E_{disl} = -J.
\end{equation}

The corresponding type of dislocations also exists for the phase
with diagonal stripes, see right panel of Fig.~\ref{Fig16}. Again,
calculating the energy of four  spins around point A in the right
panel of Fig.~\ref{Fig16}, we find that the energy of the
dislocation equal to that for the dislocation shown in the left
panel of Fig.~\ref{Fig16}

\begin{equation}\label{eq:22}
 E'_{disl} = -J.
\end{equation}

\begin{figure} \centering
\includegraphics[width=0.8\textwidth]{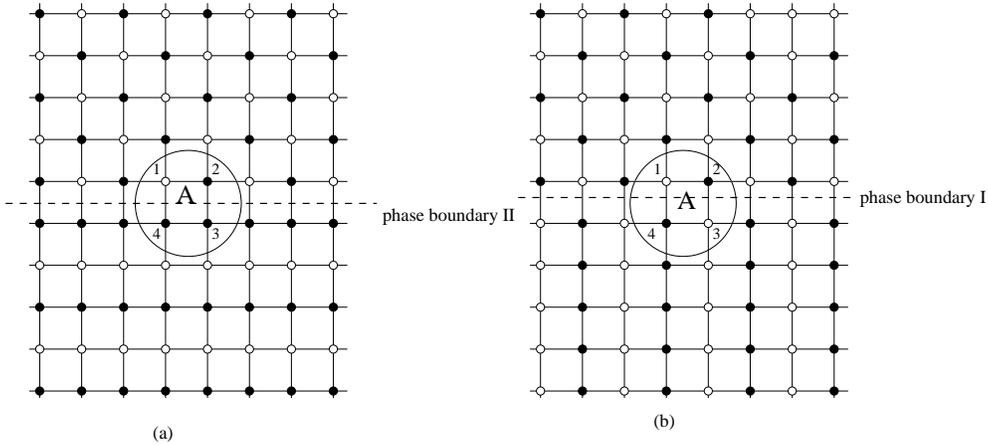}
\caption{Boundaries between phases with with diagonal stripes and
 (\emph{a}) horizontal and (\emph{b}) vertical stripes.}
 \label{Fig17}
\end{figure}

Now, let us consider the energy of phase boundaries between phases
with horizontal/vertical and diagonal stripes. Two types of such a
boundary  are illustrated in Fig.~\ref{Fig17}.

Following the same procedure  of calculating the energies of four
sites around point A both in panels (\emph{a}) and (\emph{b}) in
Fig. \ref{Fig17}, we find the values of energy per site
corresponding to phase boundaries I and II. Thus, we have

\begin{eqnarray}\label{eq:23}
E_{pbI} &=& -\frac{3}{2}J + J',\nonumber\\
E_{pbII} &=& 0. \end{eqnarray}

\begin{figure}[h] \centering
\includegraphics[width=0.8\textwidth]{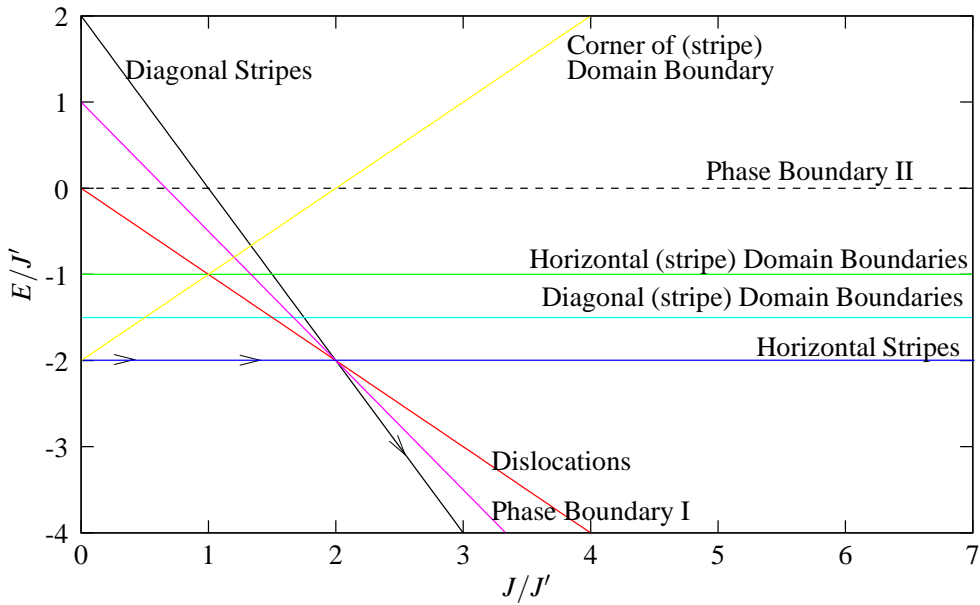}
\caption{(Color online) Energies (per site) of different phases
and  main topological defects for the square Ising lattice with
nearest-neighbor and diagonal interactions versus parameter
$\alpha = J/J'$. The solid line with arrows illustrates the
variation of parameters corresponding to the adiabatic
computations, similarly to that in Fig.~\ref{Fig8}.} \label{Fig18}
\end{figure}

The energies of all phases and defects per site are summarized in
Fig.~\ref{Fig18}.

From Fig.~\ref{Fig18}, we can see that the crossing point of
energies for phases with diagonal and horizontal/vertical stripes
($J/J' = 2)$ is also the crossing point for the energies of
dislocations. On the other hand, the entropy of a state with the
defects is significantly larger than the entropy of a plain
disordered state. This implies, of course, the possibility of the
creation of dislocations near the crossover between these two
phases. Therefore, at nonzero low temperatures, the defects
described above will proliferate into the ordered
antiferromagnetic checkerboard and stripy states. The
proliferation phenomenon occurs within a broad range of $J/J'$
values. Due to this proliferation of topological defects, there
arises a problem with performing AQC. It has the same origin as we
have discussed for the five-$\pi$-ring plaquette. Namely, during
adiabatic evolution of the ground state, i.e. the
antiferromagnetic checkerboard state, when the ratio $J/J'$
decreases,  the system will enter into the intermediate phase
characterized by a creation of a large number different
topological defects. Such topological defects are quite stable and
may have a very long lifetime provided that they do not annihilate
with each other like vortices with antivortices. As the result of
this adiabatic evolution a some number of defects will definitely
proliferate into the final ground state and therewith may
introduce an error in the adiabatic quantum computation. At
present, we do not know, how to resolve this issue of the
spontaneous creation of topological defects during the AQCs with
the larger clusters. Note that even for the case of the planar
array of the $\pi$-rings, when the coupling is not tuned and the
value  $\alpha= 2\sqrt{2}$, we expect that the proliferation of
the defects will strongly modify the ordered antiferromagnetic
state. The similar possibility was also revealed in the analysis
of thermodynamics of a 5-site plaquette.

\section{Adiabatic quantum computations  with a planar array of
$\pi$-rings and conclusion}

\begin{figure}[htp]\centering
   \subfigure[$t = T/J' = 0.1$]{
      \includegraphics[scale=0.25]{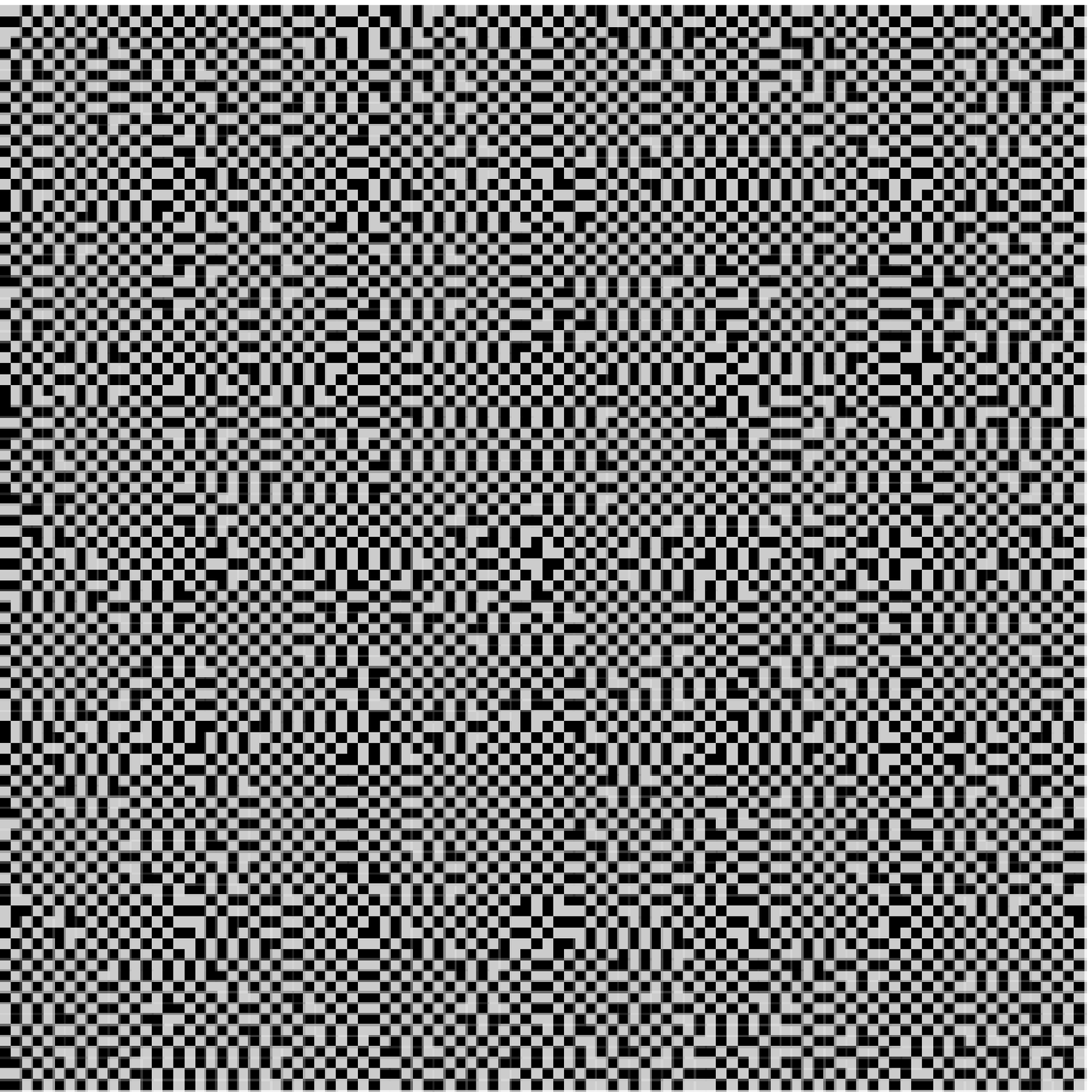}}
   \subfigure[$t = T/J' = 0.25$]{
      \includegraphics[scale=0.25]{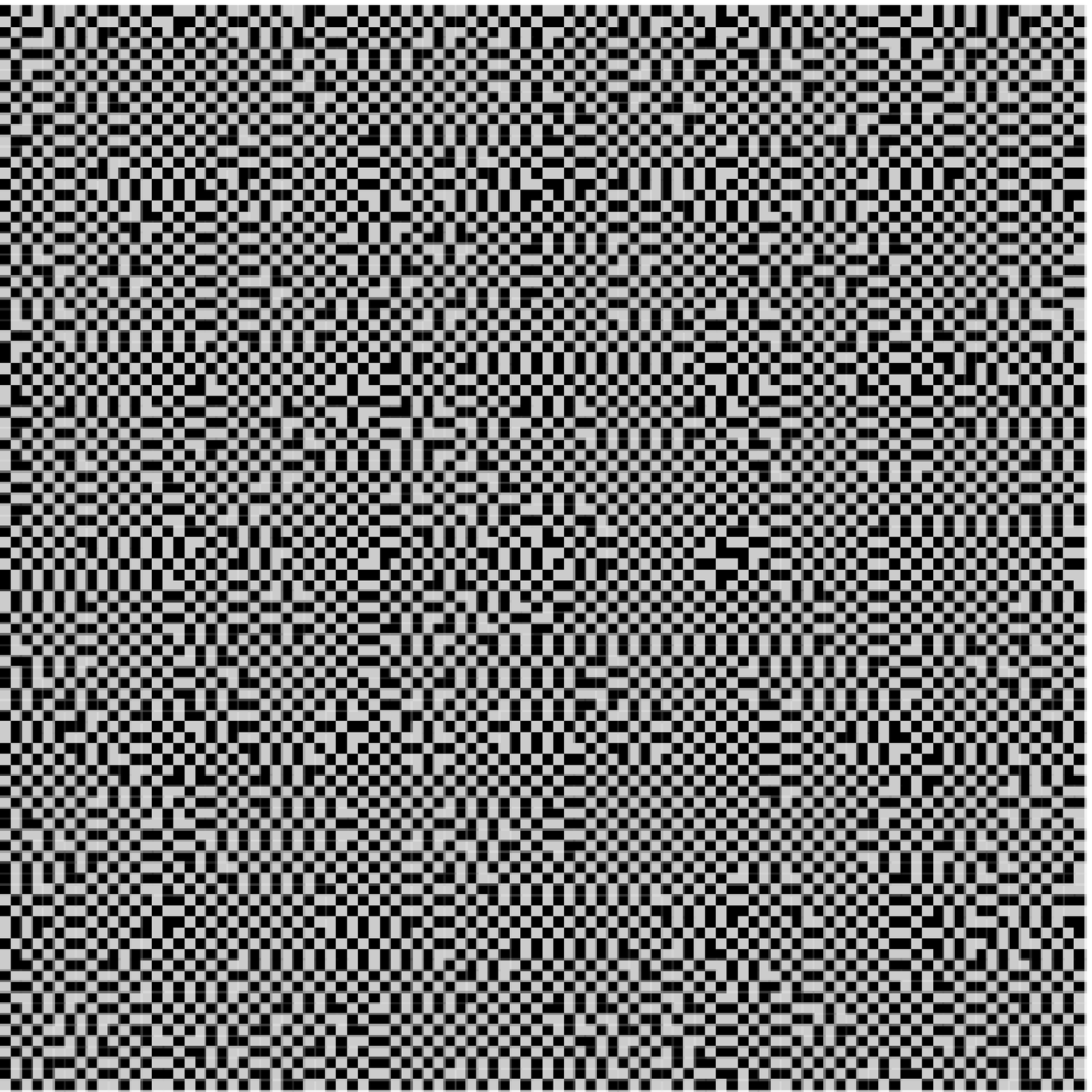}}
   \subfigure[$t = T/J' = 1.0$]{
      \includegraphics[scale=0.25]{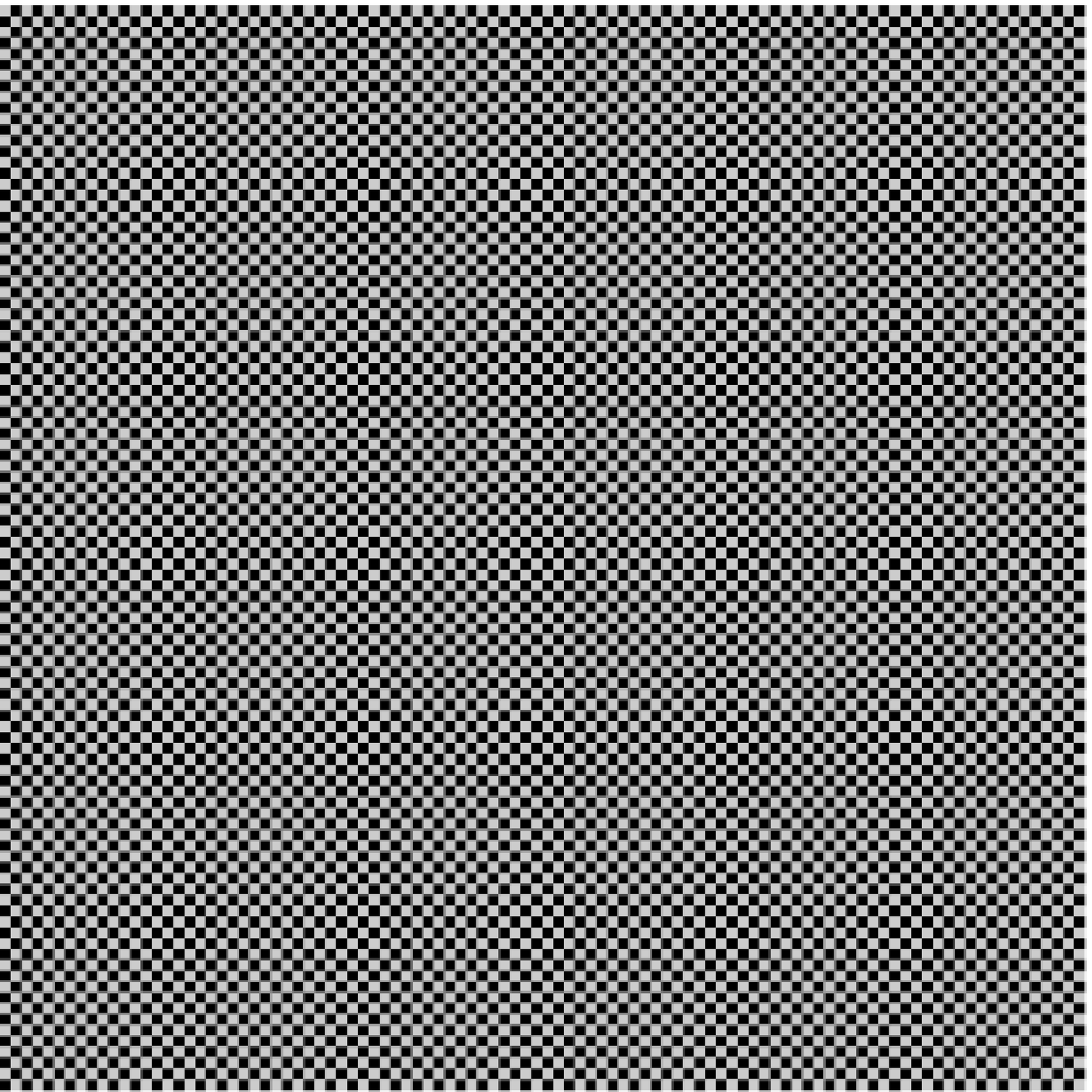}}
   \subfigure[$t = T/J' = 2.5$]{
      \includegraphics[scale=0.25]{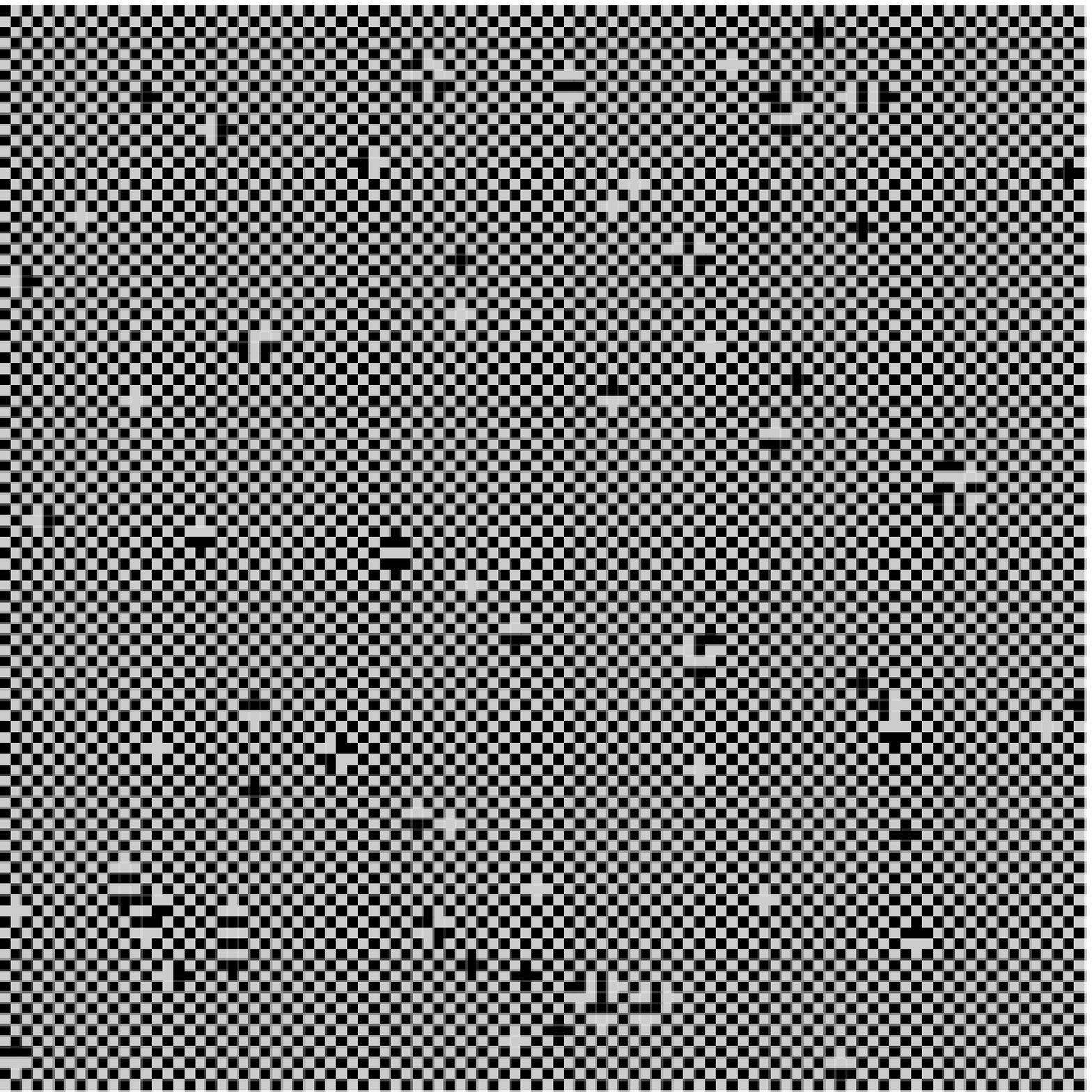}}
   \subfigure[$t = T/J' = 3.0$]{
      \includegraphics[scale=0.25]{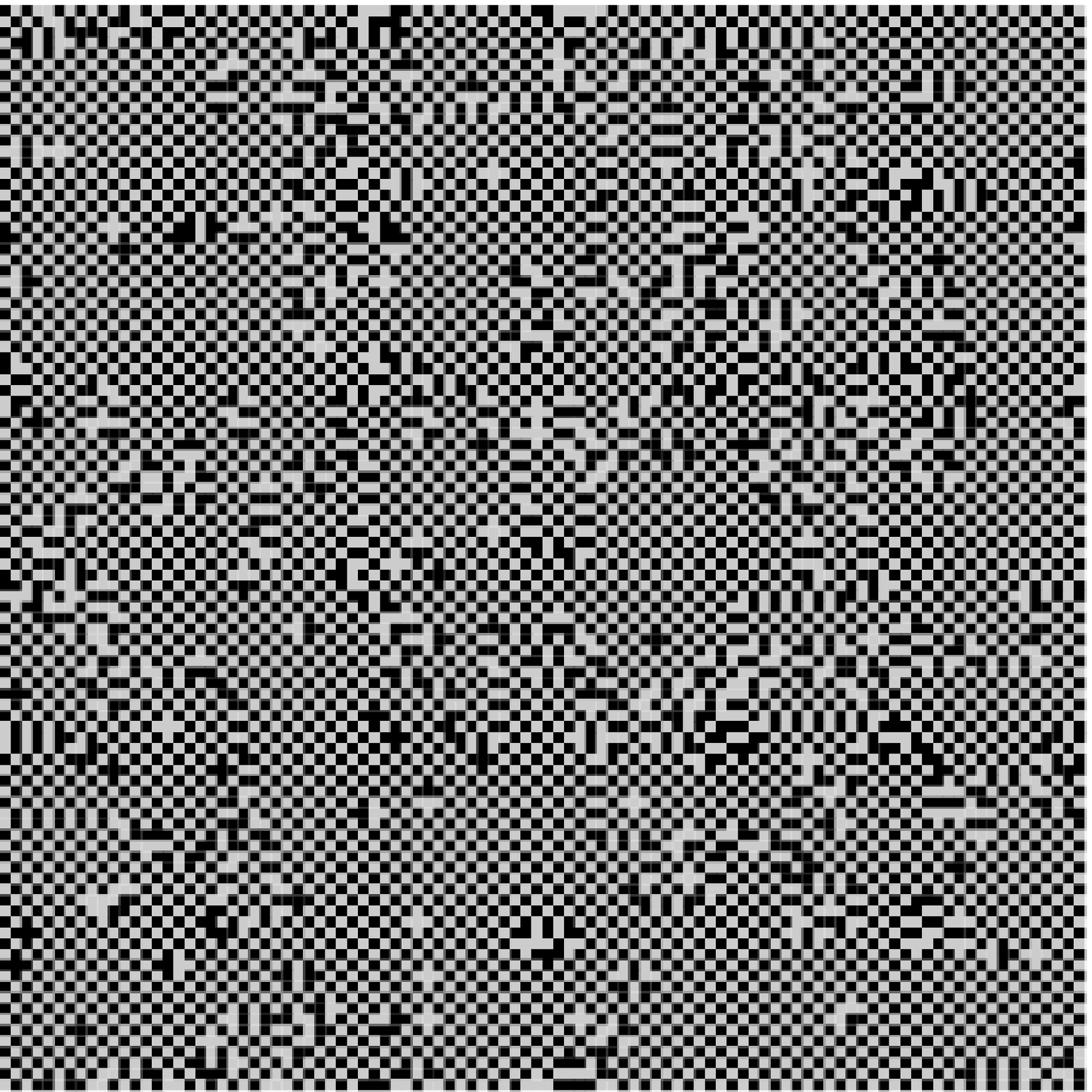}}
   \subfigure[$t = T/J' = 4.0$]{
      \includegraphics[scale=0.25]{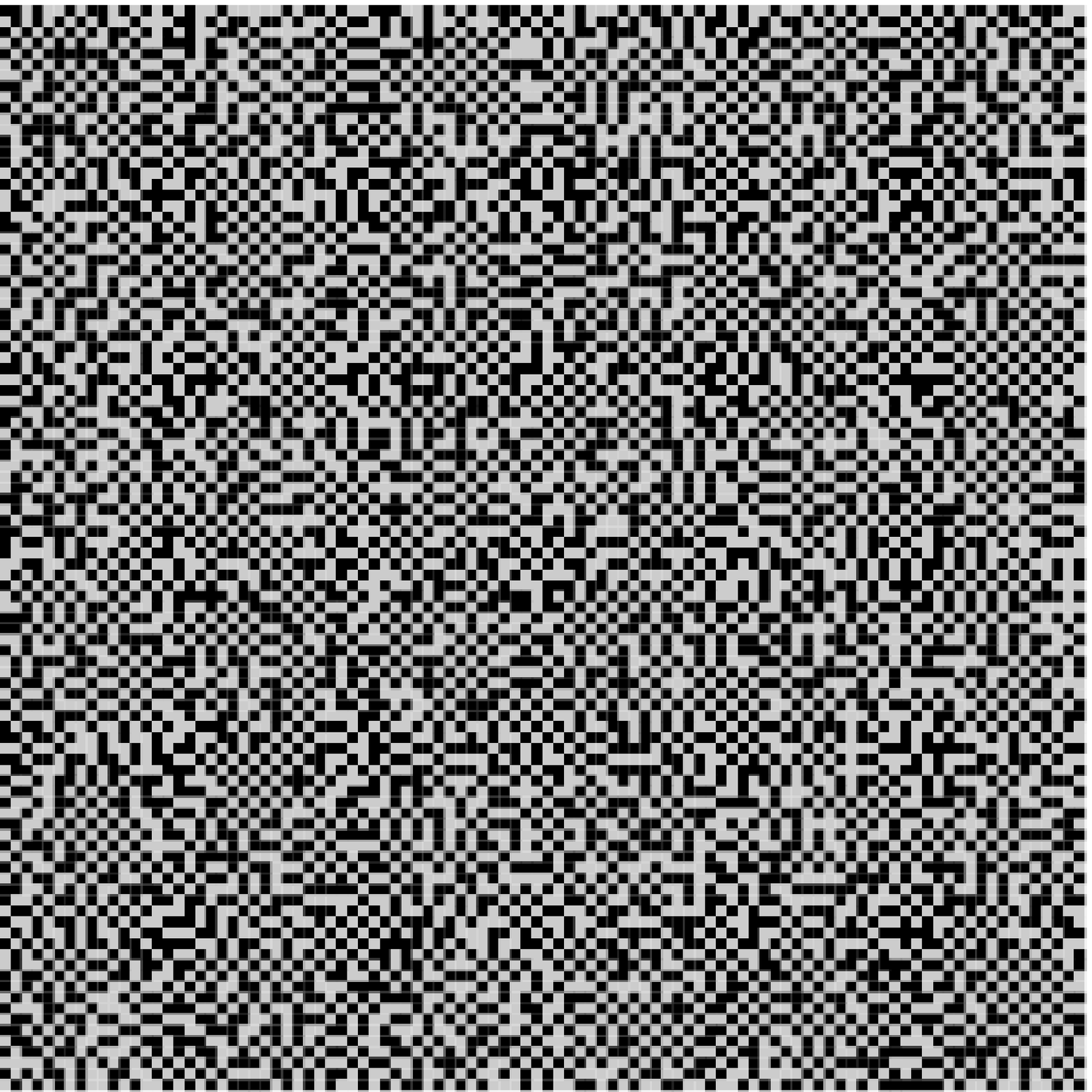}}
\caption{The result of Monte Carlo simulations of the spin
structure for the $100\times100$ square Ising lattice with
competing antiferromagnetic interactions at $\alpha = J/J' =
2\sqrt{2}$. Dark and light square cells correspond to $s = +1$ and
$s = -1$, respectively.}
\label{Fig19}
\end{figure}

Now, it is interesting to consider a realistic system, where the
physical phenomenon of the proliferation of defects described
above can be realized.  The first choice is the planar array  made
of $\pi$-rings. We would like to consider a large system like a
$100\times100$ square lattice with Ising spins having competing
antiferromagnetic interactions at arbitrary value of $\alpha =
J/J' $. The specific case of the dipole-dipole interaction between
isolated $\pi$-rings deposited onto a dielectric substrate will
correspond to the value $ \alpha=2\sqrt{2}$. We performed the
numerical simulation of this system using the Monte Carlo method,
which is the only method available to treat such large systems.
Let us demonstrate the results of the Monte Carlo simulation at
various temperatures ($T/J'=0.01 - 5.0$). The spin structure for
the $100\times100$ square Ising lattice with competing
antiferromagnetic interactions at $\alpha = J/J' = 2\sqrt{2}$ is
presented in Fig.~\ref{Fig19}, where each square in this figure
corresponds to the $\pi$-ring. The dark square corresponds to the
up-spin orientation of the orbital moment, while the light square
corresponds to the down-spin orientation of the orbital moment. At
this value of $\alpha=2\sqrt{2}$ corresponding to the square array
of $\pi$-rings (see, the Ref.~\cite{KirtleyPRB05}), at low
temperatures we should expect a usual two-sublattice
antiferromagnetic ordering of Ising spins. Instead, we see a
complicated array, a mixture of light and dark squares. The
two-sublattice portions are mixed with the topological defects
like dislocations and domain boundaries. This picture is in the
agreement with our previous discussion demonstrating the
possibility of a rather low energy barrier and large entropy for
the  creation of defects. A similar picture of the spin
distribution was also observed in experiments with the arrays of
$\pi$-rings \cite{KirtleyPRB05}.

Let us discuss the possibility to use a large planar arrays of
$\pi$-rings for adiabatic quantum computations. The
semiqualitative discussion presented in this paper clearly
demonstrates that the clusters of $\pi$-rings may be perfectly
described by the Ising model with competing nearest-neighbor and
diagonal interactions. The system exhibits a plethora of unusual
properties, which are promising for the analysis of various
analogous physical systems of different nature and also for
studies of different AQC algorithms. However, as we discussed
above, the proliferation of topological defects arising in the
process of AQC may contribute an error into the final result.  The
physical phenomenon of the proliferation of defects described for
the planar array made of $\pi$-rings may also lead to a formation
of a new type of glassy state~\cite{OHare}.

Recently, a scalable design for adiabatic quantum computations has
been proposed and realized~\cite{Zakosarenko}. The key element of
this design is a coupler, which is a ring with a single Josephson
junction that provides a controllable coupling between two
bistable flux qubits. The similar tunable coupling may be realized
for the large  $\pi$-ring arrays, which we discuss here. Now let
us investigate how good is to use such $\pi$-ring arrays with
controllable coupling for adiabatic quantum computations. Such
arrays are well described by the Ising model with competing
interactions as we discussed in the paper. The adiabatic quantum
computer based on such large $\pi$-ring arrays with the
controllable coupling will be able to solve a very limited range
of problems. In many cases they are still toy problems, which
serve as a polygon towards future developments for various
realistic applications. One of such applications is a Travelling
Salesman Problem~\cite{Kieu}, which can be represented in the form
of more complicated Ising model, with a set of a coupling
constants~\cite{Tosatti1}.

Now let us investigate  the process of  the adiabatic evolution of
$100 \times 100$  $\pi$-ring array when the coupling between all
$\pi$-rings is tuned simultaneously in a way that the ratio
$\alpha=J/J'$, which is the ratio of exchange constants in the
corresponding Ising model, increases from zero. First, we consider
zero or very close to zero temperatures.  When the value of
$\alpha$ is small  we have a well defined ground state -- the
stripe phase. The distribution of spins for such a state form
stripes oriented horizontally or vertically. So, it is double
degenerate and, therefore, there is a possibility to form two
types of domains associated with horizontal and vertical stripe
phases as discussed in the previous sections. When the parameter
$\alpha$ increases adiabatically according to the adiabatic
theorem (see, a detail discussion of the application of this
theorem to quantum adiabatic computations given in Ref.
\cite{Kieu1}) the system will remain in the ground state. For
small temperature the domain walls and domains may still
proliferate into the system but if they do, their number is very
small.  When the parameter $\alpha$ approaches to the value
$\alpha = 1$, the proliferation of numerous domain walls and other
types of topological defects such as dislocations and phase
boundaries described in the previous section should arise. Within
the $1 < \alpha< 2$ range, the state is highly disordered. The
formation of such disordered structures in the $\pi$-ring arrays
arising during the adiabatic evolution in the course of tuning the
coupling constants may be explained by the presence of strong
frustrations in this system. The frustration reaches its maximum
in the vicinity of the value $\alpha=2$.

 \begin{figure}[htp]\centering
   \subfigure[$\alpha = J/J'= 1.0$]{
      \includegraphics[scale=0.25]{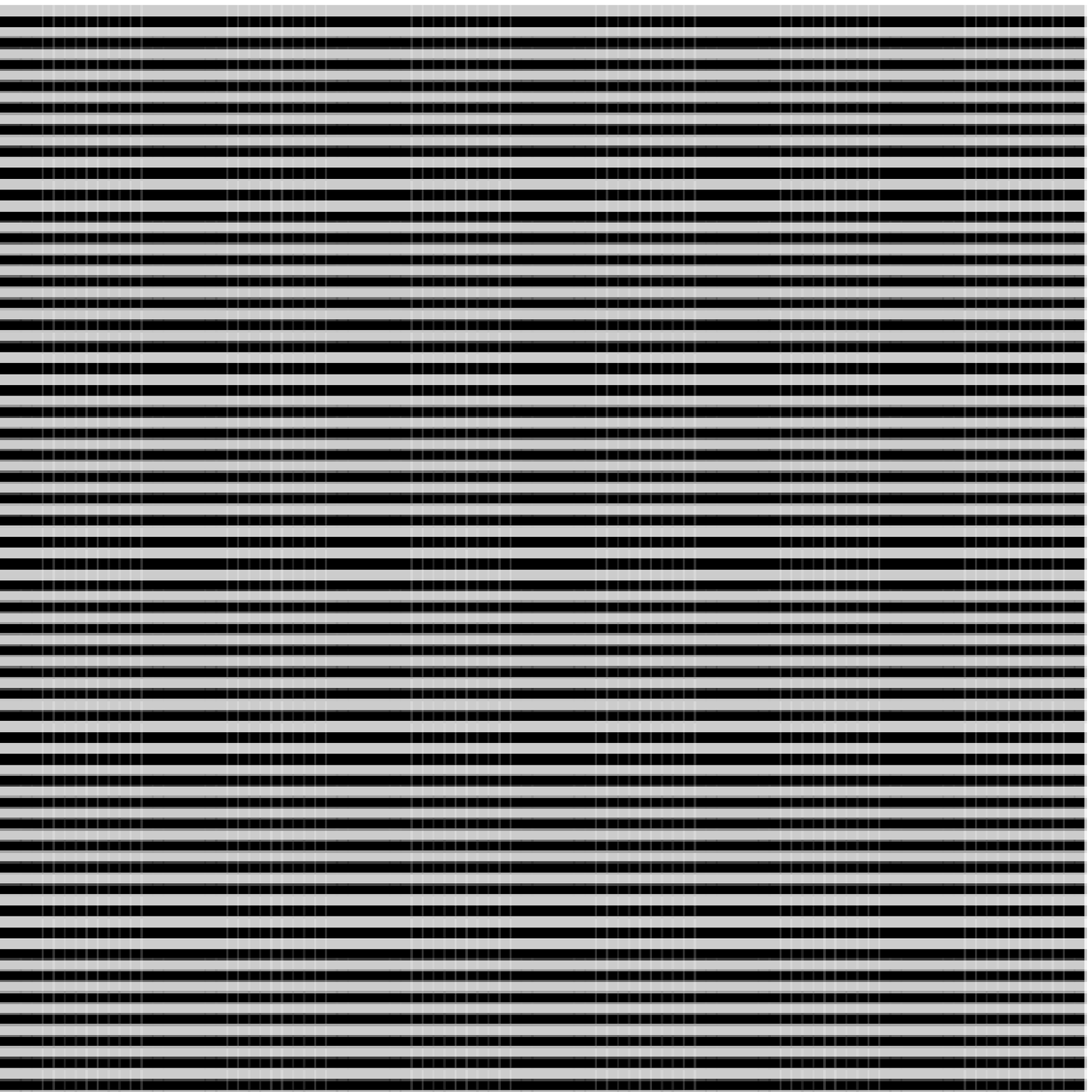}}
   \subfigure[$\alpha = J/J'= 1.6$]{
      \includegraphics[scale=0.25]{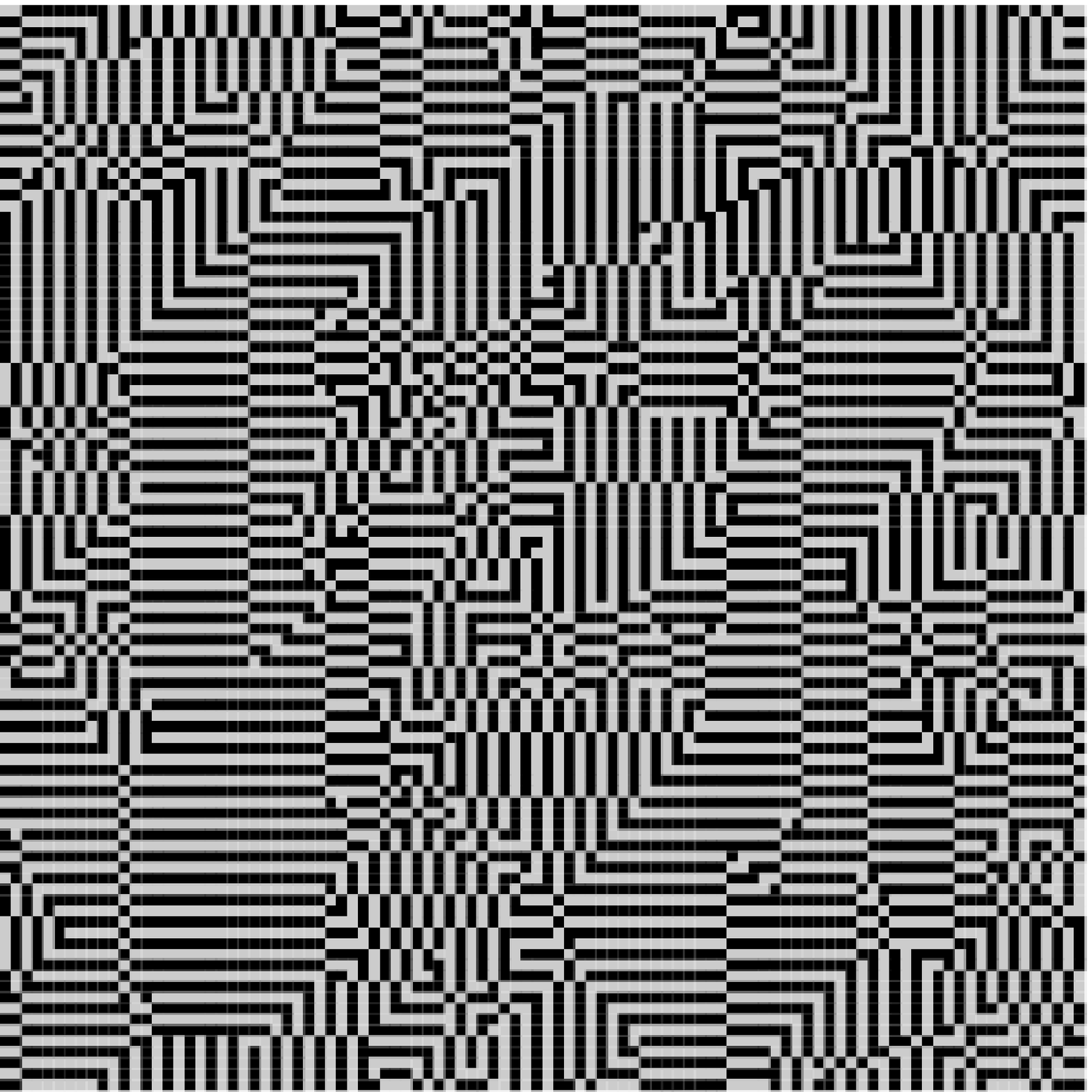}}
   \subfigure[$\alpha = J/J'= 1.8$]{
      \includegraphics[scale=0.25]{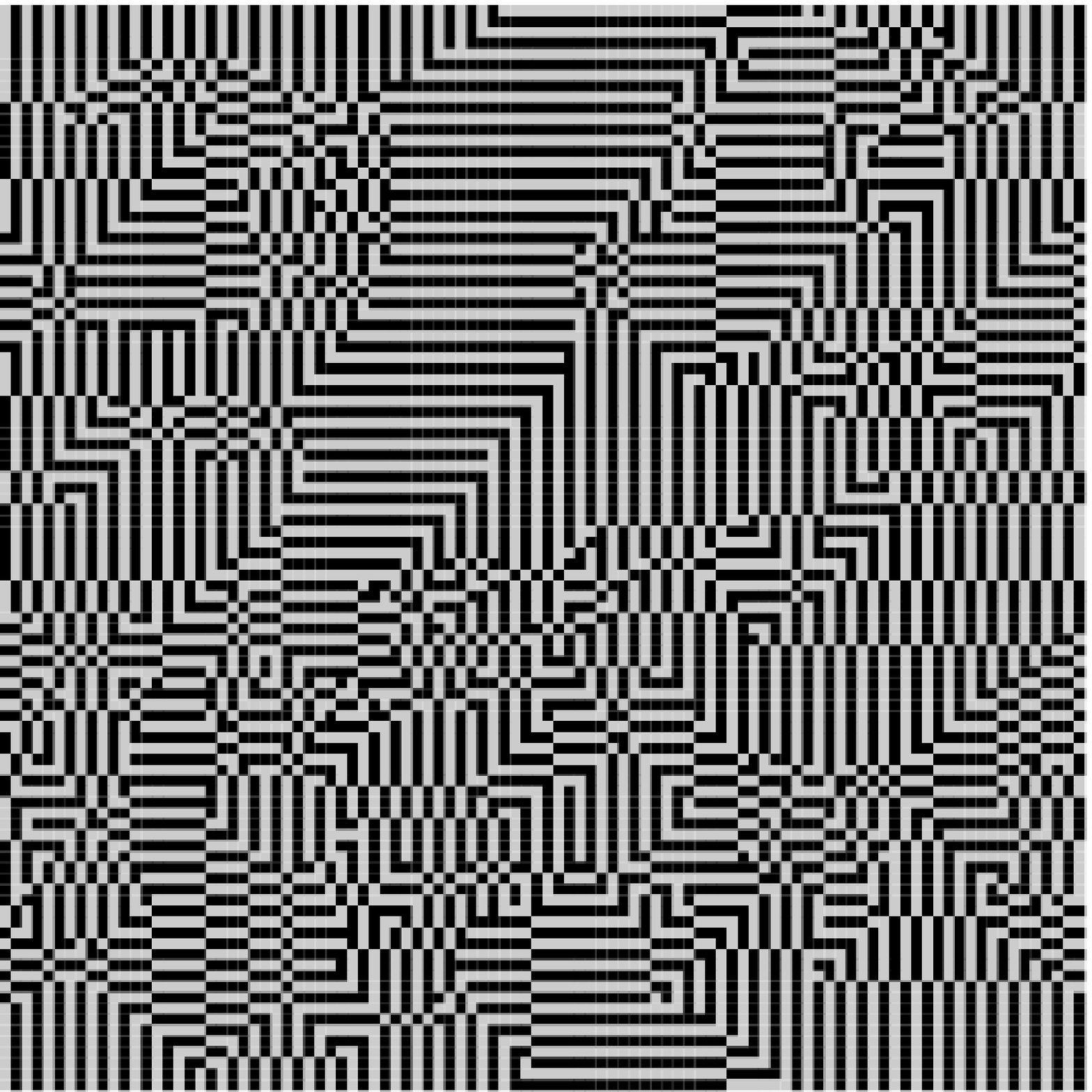}}
   \subfigure[$\alpha = J/J'= 2.0$]{
      \includegraphics[scale=0.25]{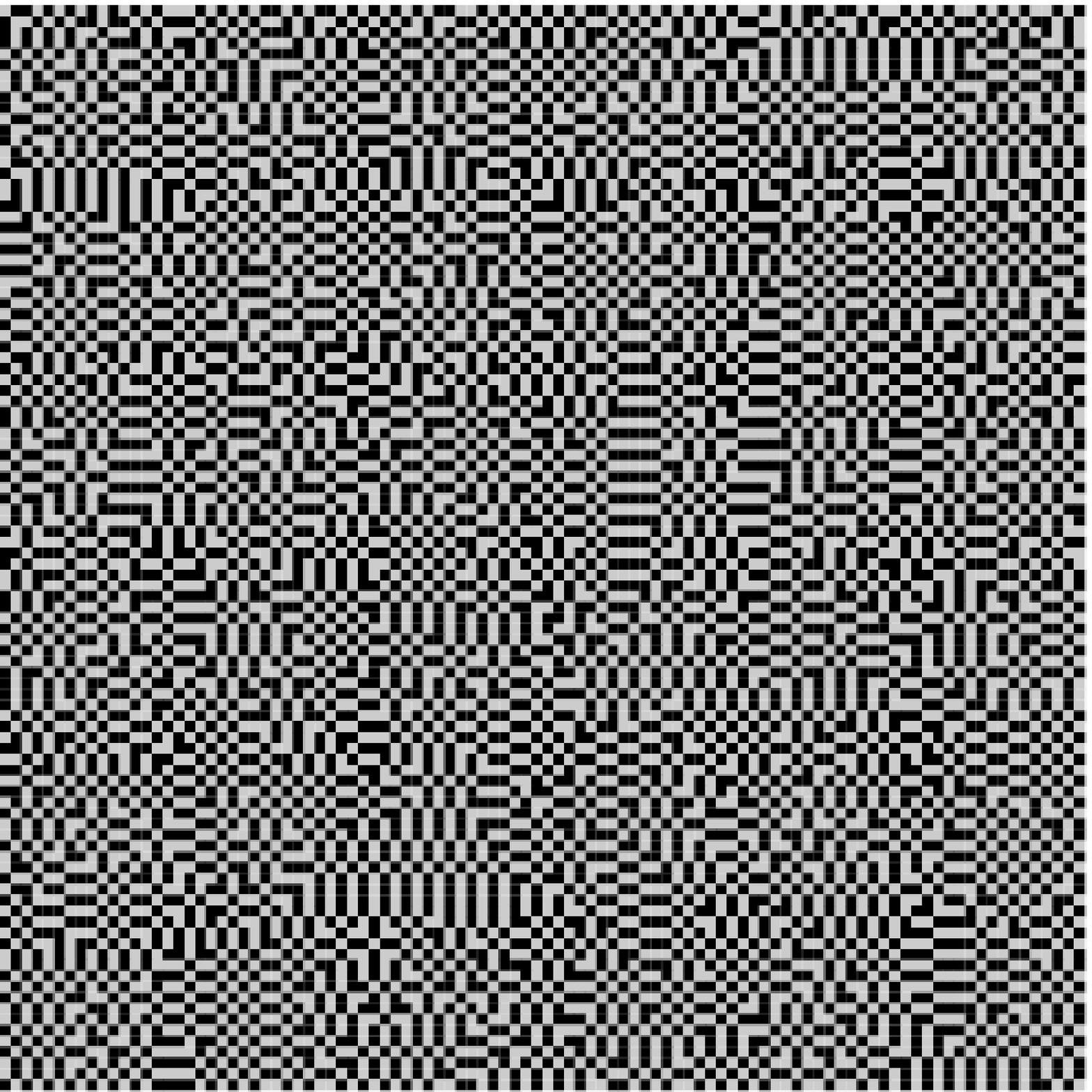}}
   \subfigure[$\alpha = J/J'= 2.4$]{
      \includegraphics[scale=0.25]{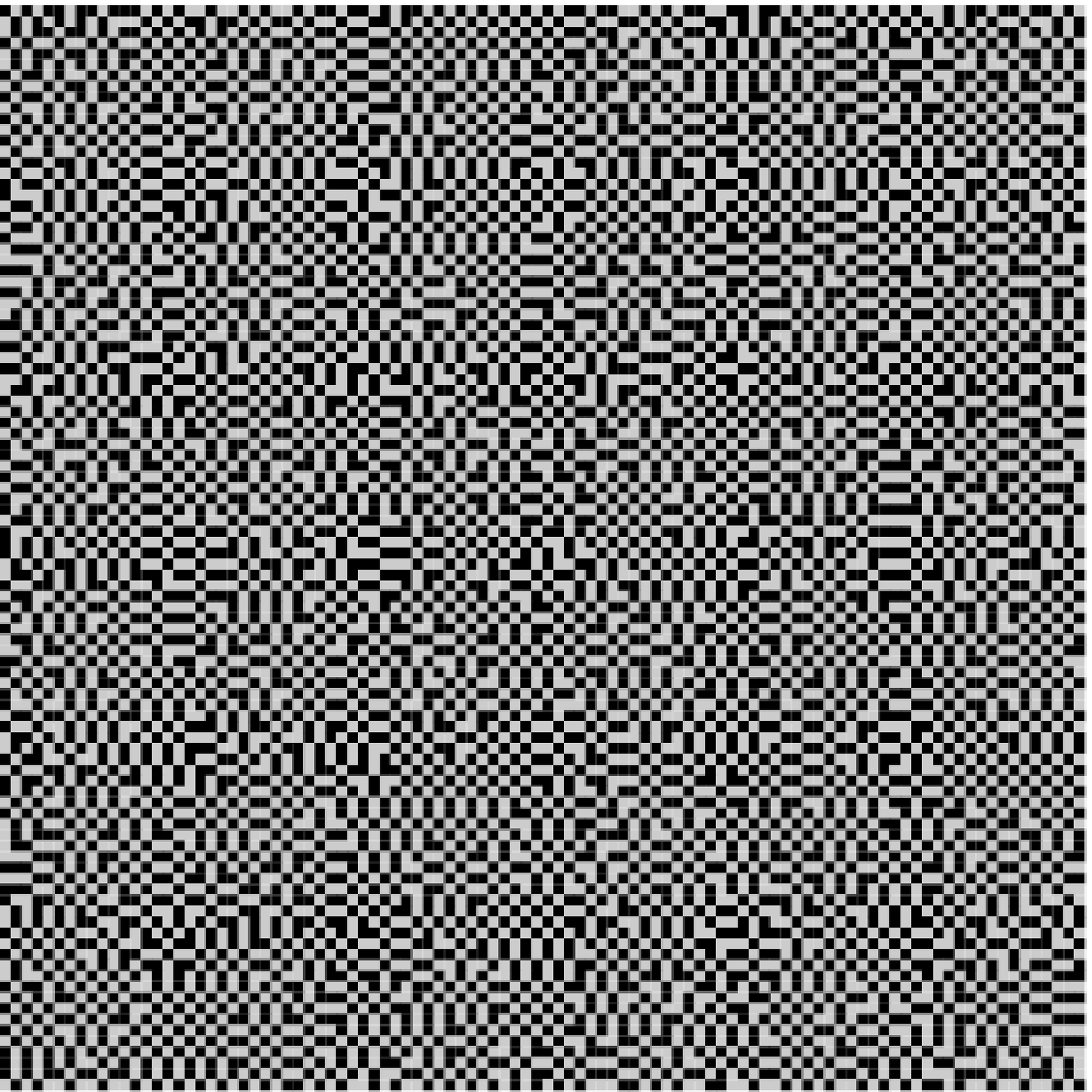}}
   \subfigure[$\alpha = J/J'= 3.4$]{
      \includegraphics[scale=0.25]{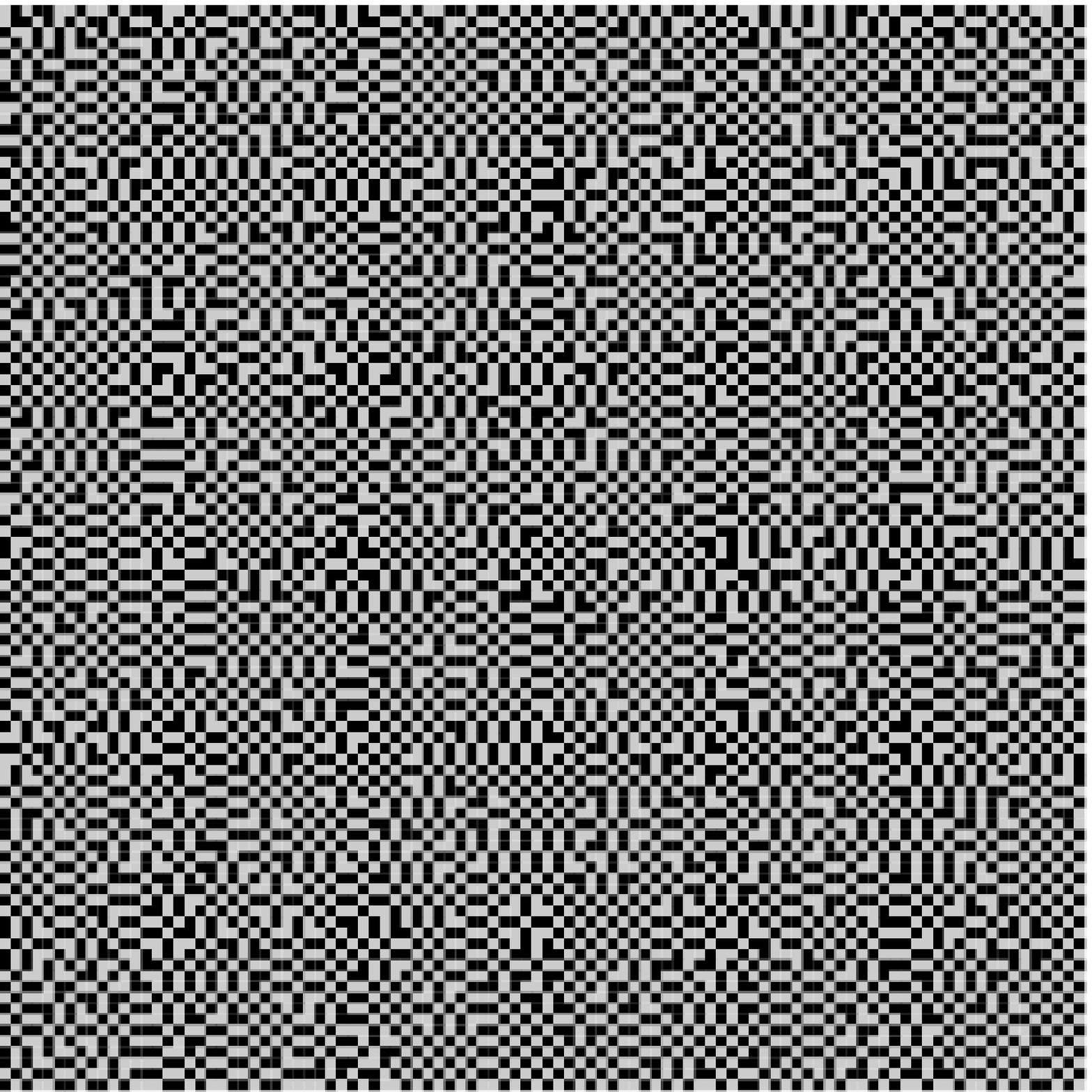}}
\caption{The result of Monte Carlo simulations of the spin
structure for the $100\times100$ square Ising lattice with
competing antiferromagnetic interactions at different $\alpha =
J/J' $ and at fixed low temperature $T/J'=0.1$. Dark and light
square cells correspond to $s = +1$ and $s = -1$, respectively.}
\label{Fig20}
\end{figure}

 \begin{figure}[htp]\centering
   \subfigure[$\alpha = J/J'= 1.0$]{
      \includegraphics[scale=0.25]{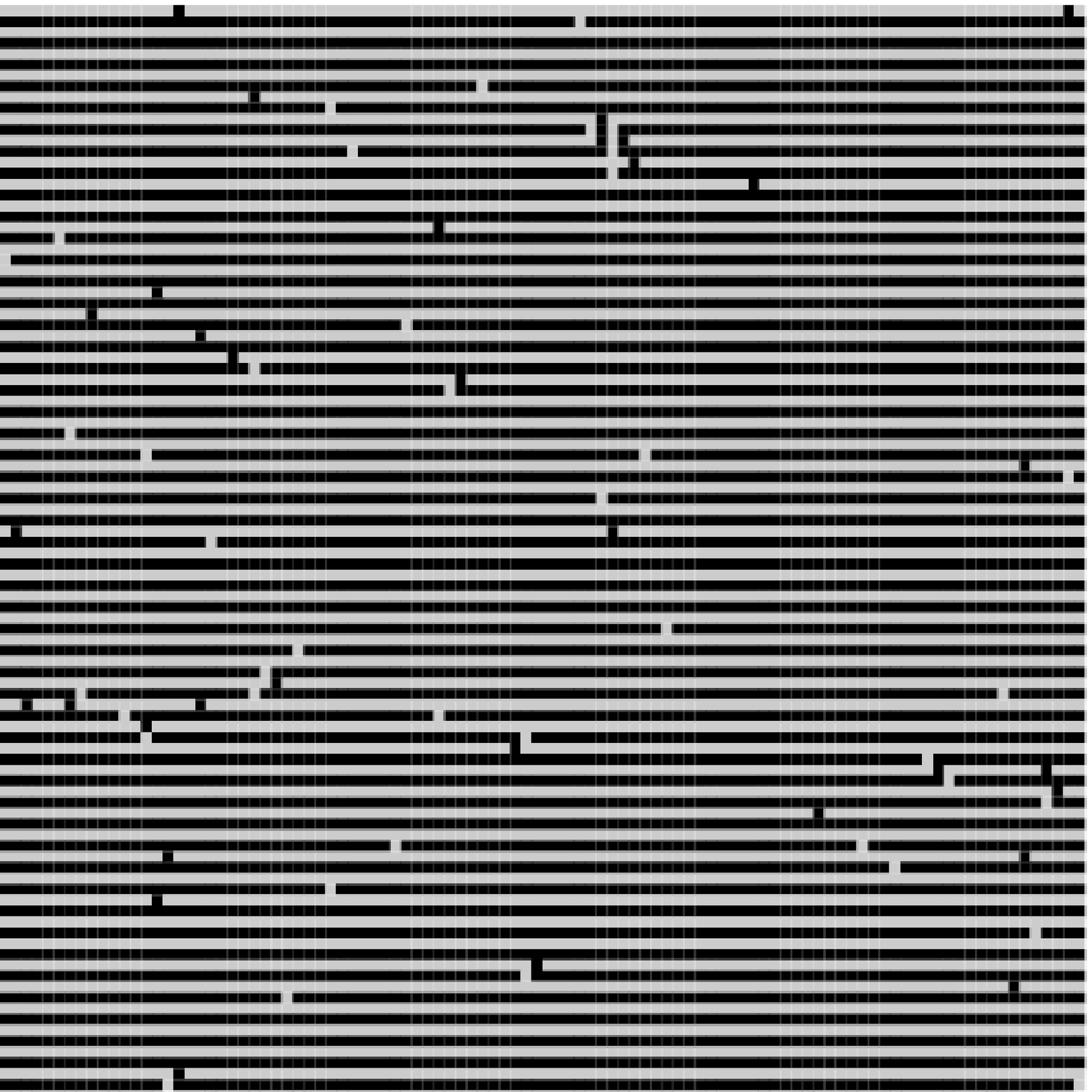}}
   \subfigure[$\alpha = J/J'= 1.6$]{
      \includegraphics[scale=0.25]{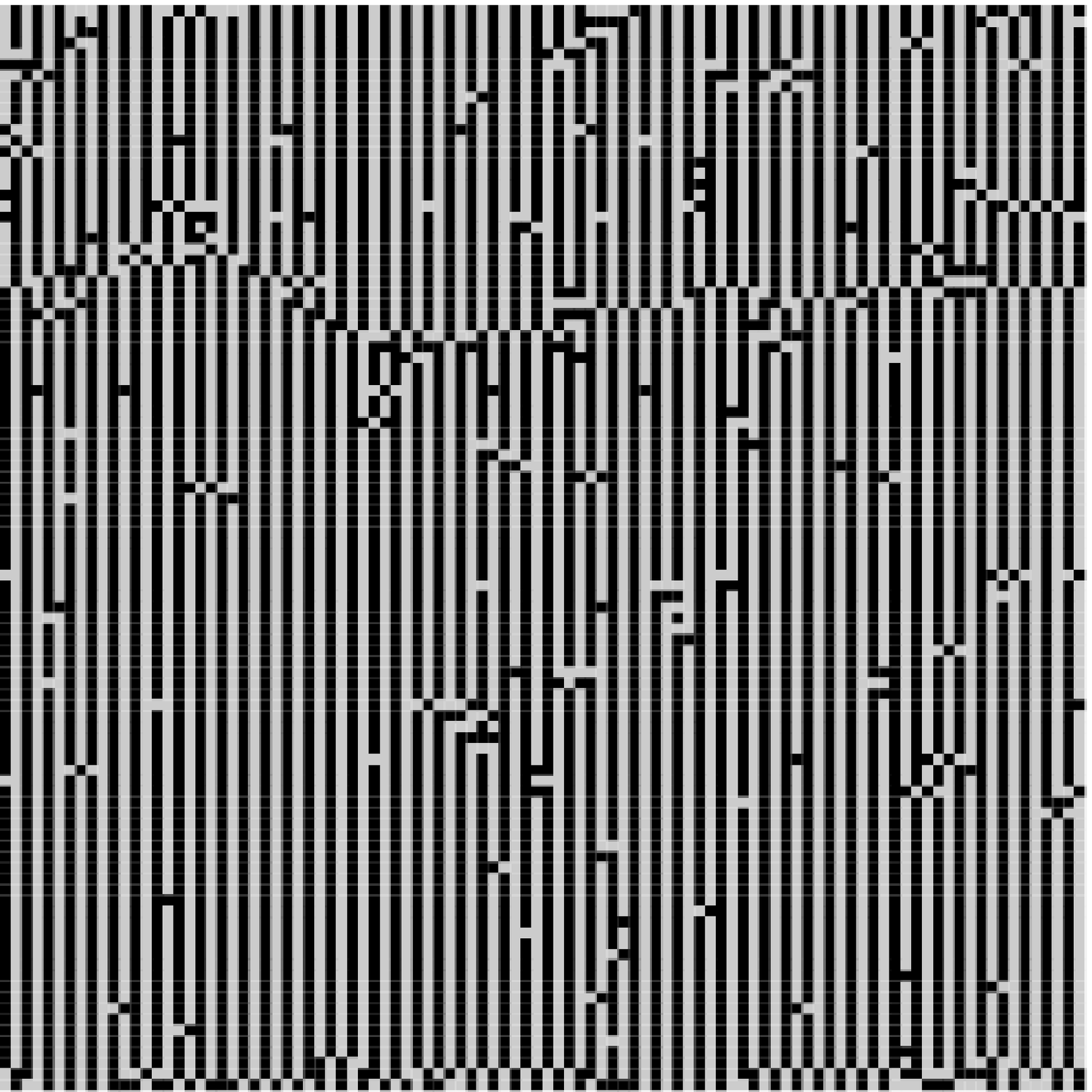}}
   \subfigure[$\alpha = J/J'= 1.8$]{
      \includegraphics[scale=0.25]{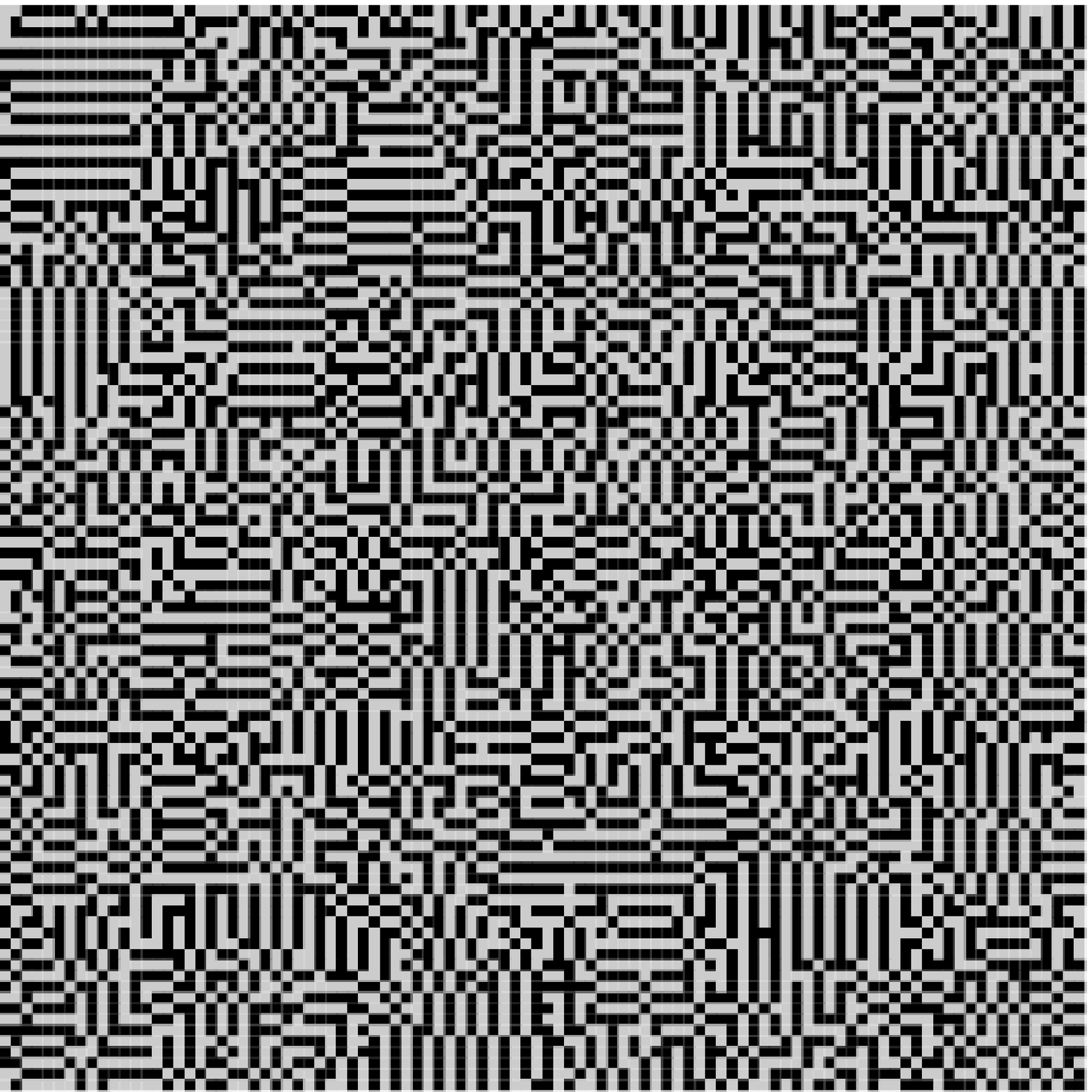}}
   \subfigure[$\alpha = J/J'= 2.0$]{
      \includegraphics[scale=0.25]{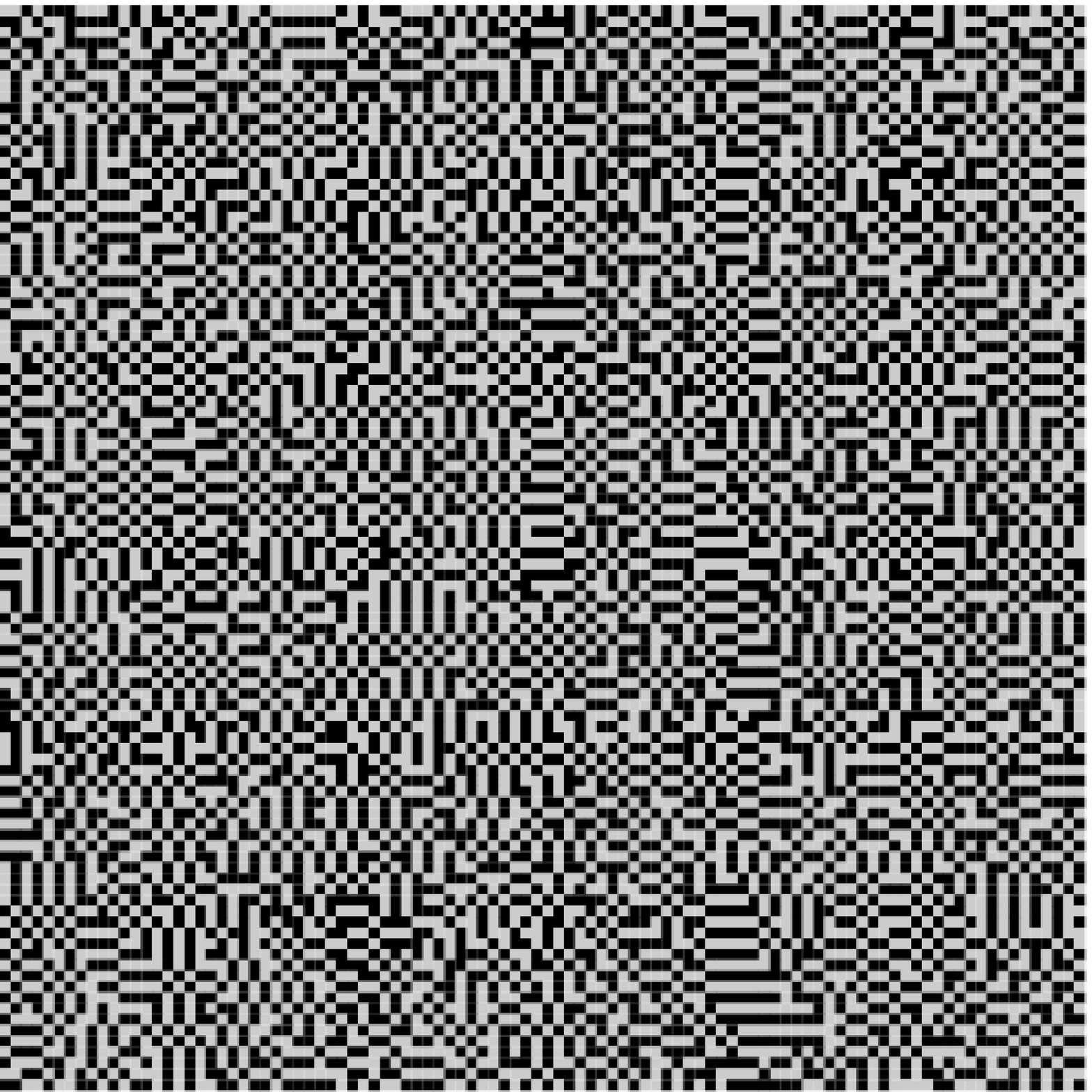}}
   \subfigure[$\alpha = J/J'= 2.4$]{
      \includegraphics[scale=0.25]{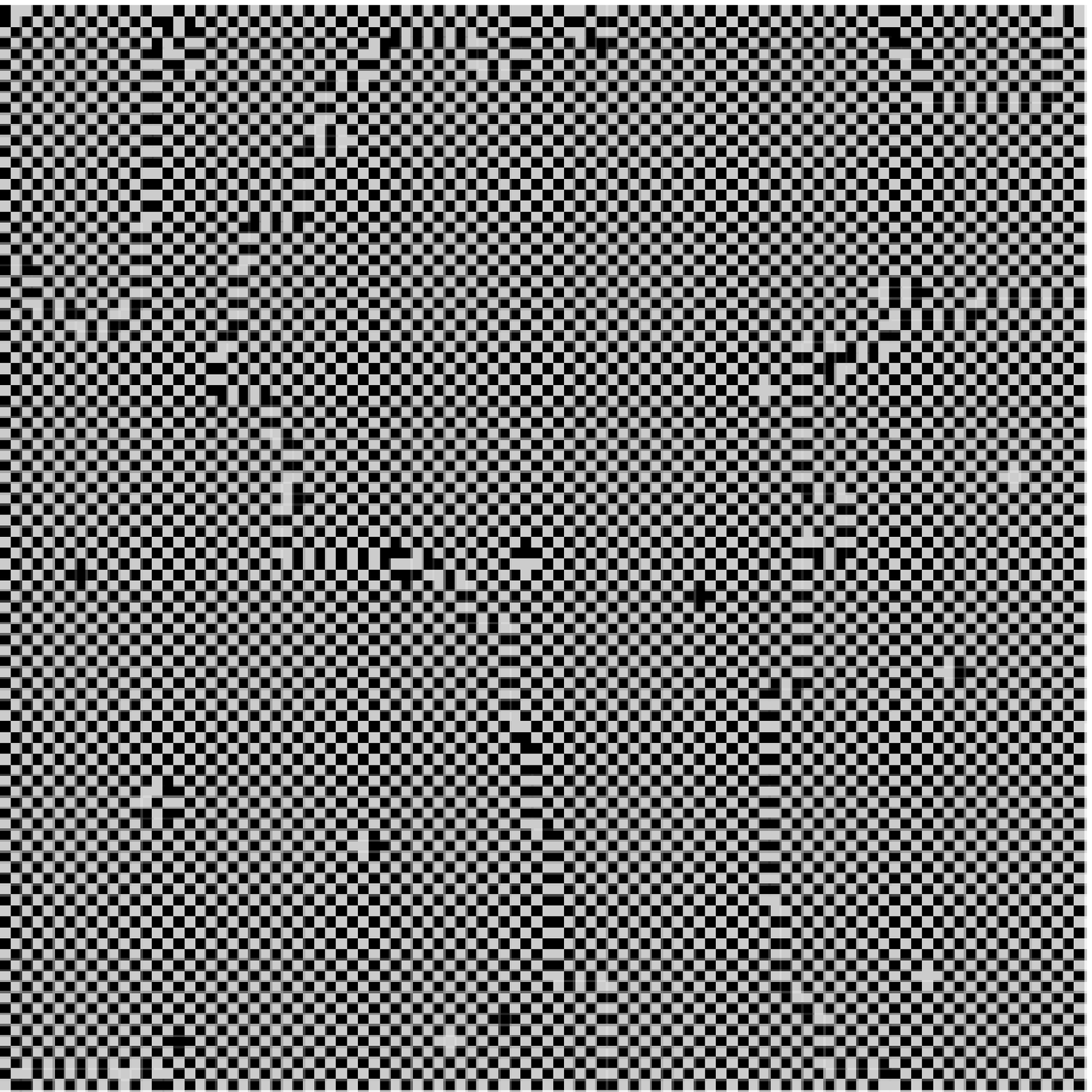}}
   \subfigure[$\alpha = J/J'= 3.4$]{
      \includegraphics[scale=0.25]{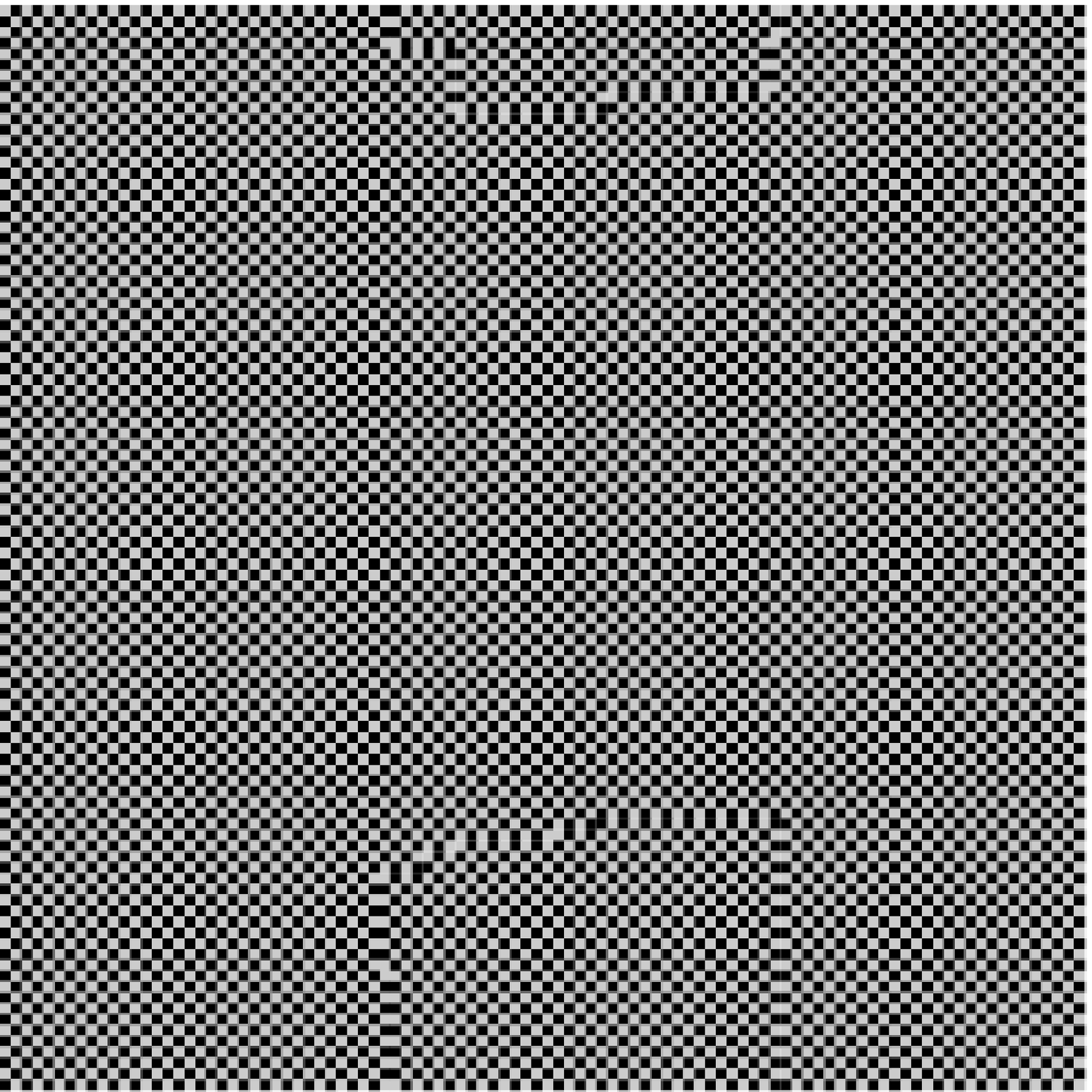}}
\caption{The result of Monte Carlo simulations of the spin
structure for the $100\times100$ square Ising lattice with
competing antiferromagnetic interactions at different $\alpha =
J/J' $ and at fixed temperature $T/J'=1.5$. Dark and light square
cells correspond to $s = +1$ and $s = -1$, respectively.}
\label{Fig21}
\end{figure}

\begin{figure}[htb]\centering
\includegraphics[scale=0.6]{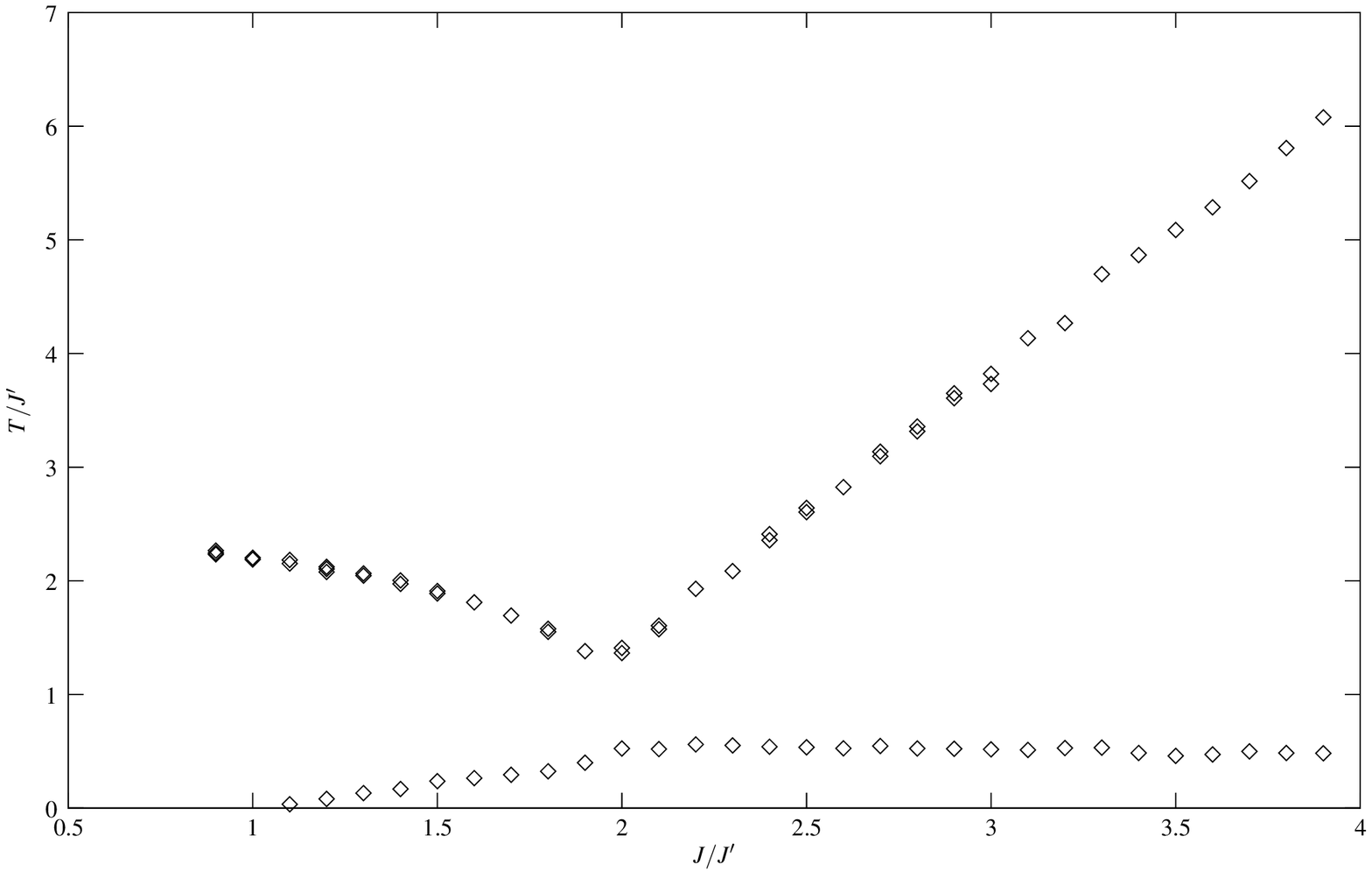}
\caption{A schematic phase diagram the 2D Ising model with
competing interactions on a square lattice based on the Monte
Carlo calculations of heat capacity for the $100 \times 100$
square plaquette. The lower branch corresponds to the boundary of
the glassy state and the upper branch is the boundary of the
phases with the long-range spin order.}
\label{Fig22}
\end{figure}

Our estimation indicate that in this $\alpha$ parameter range the
energies of many topological defects, like domain walls  between
different antiferromagnetic phases as well as various
dislocations, are very close to the ground state energy.
Therefore, there appear many locally stable (or metastable) states
associated with local energy minima separated by  energy barriers
(see, Fig.~\ref{Fig20}d). With the further increase of $\alpha$,
the energy barriers separating these metastable states increase.
During the adiabatic evolution the system may be trapped into one
of these metastable minima. After that, during the subsequent
adiabatic evolution, it will have not enough time to leave this
metastable minimum and and therefore the system will remain in the
disordered state associated with this minimum. In fact, during
this mass proliferation of topological defects, we enter into the
disordered glassy state~\cite{OHare,Tosatti}. The glassy state
consists of many equivalent minima. Obviously during the adiabatic
computing the system will be trapped in one of them. If we repeat
the adiabatic computing once more, the system may be trapped into
another equivalent minimum of the glassy state. Performing this
AQC many time a complicated glassy state in a form of a large
number of disordered ground states associated with closed energies
may be reproduced. For $\pi$-ring clusters, there is a
straightforward way to read out the final result in the form of
the configuration of the orbital moments, as it was done in the
Ref.~\cite{KirtleyNature}. The adiabatic computing processes
described here may lead to the same results as in the case of the
thermal relaxation but can be sometimes significantly faster, see
e.g. Ref.~\cite{Zagoskin}.

Thus, we see that the adiabatic evolution of the system may lead
to an efficient adiabatic computation corresponding to a very
complicated ground state, which can be characterized by the
Edwards-Anderson order parameter~\cite{EdwAnd}. The final result
of this evolution is quite nontrivial: a glassy state associated
with frustrations.  It is also very interesting to investigate,
how such adiabatic evolution will depend on temperature,
especially taking into account that the thermal environment can
enhance the performance of adiabatic
computations~\cite{TherAssAQC}. Note here that an increase in
temperature could help one to avoid trapping the system into a
complicated glassy state. This situation is illustrated in
Fig.~\ref{Fig21}. It corresponds to the same variation of the
$J/J'$ parameter as in Fig.~\ref{Fig20}, but at higher temperature
$T/J' = 1.5$. We see that in contrast to the low-temperature case,
it is possible to pass from the stripe phase to the checkerboard
antiferromagnetic phase without being trapped in the glassy state.
Here, only in the vicinity of $J/J'= 2$, the system exhibits a
pronounced disorder.  In general, let us note that the use of the
evolution of the physical system for the computing is quite
powerful idea. It can be used at low temperatures, where
computations may have some quantum characters relying on a long
decoherence time. However, an adiabatic evolution of large
physical systems may be used for computations in a much broader
sense and at high temperatures, as it was demonstrated in
Fig.~\ref{Fig21}. We hope that these ideas will be exploited
further. The possible kinds of behavior are summarized in the
schematic phase diagram (Fig.~\ref{Fig22}) based on the Monte
Carlo calculations of the heat capacity for the $100 \times 100$
square plaquettes. This figure demonstrates that the glassy state
arising at the value $J/J'>1$ is favorable at low temperatures,
whereas in the intermediate temperature range, there exists a
crossover between different types of the long-range spin order.

A detailed description of other results for the Ising model with
competing interactions that include the Monte Carlo simulations
for the spin structure, phase diagram, and spin correlation
functions, as well as analytical and numerical results obtained by
the transfer matrix technique, will be presented in a separate
publication.

\section*{Acknowledgments}

The authors are grateful to S.~Bulgadaev, H.~Hilgenkamp,
D.~Khomskii, and J.~R.~Kirtley for helpful discussions.

The work was supported by the Royal Society (London) (grant
ISVi-2004/R2-FS), ESF network-programme AQDJJ, European network
CoMePhS, and the Russian Foundation for Basic Research (project
05-02-17600).

\end{document}